\documentclass[lettersize,journal]{IEEEtran}
\usepackage{amsmath,amsfonts,amssymb}
\usepackage[cal=esstix]{mathalpha}
\usepackage{algorithm}
\usepackage[noend]{algpseudocode}
\usepackage{array}
\usepackage[caption=false,font=normalsize,labelfont=sf,textfont=sf]{subfig}
\usepackage{textcomp}
\usepackage{bm}
\usepackage{stfloats}
\usepackage{url}
\usepackage{verbatim}
\usepackage{graphicx}
\usepackage{cite}
\usepackage{booktabs}
\usepackage{hyperref}
\usepackage{pbox}
\usepackage[dvipsnames]{xcolor}
\hypersetup{%
  colorlinks=true,%
  linkcolor={blue},
  citecolor={blue},
  urlcolor={blue},
  bookmarksnumbered=true,%
  bookmarksopen=true}
\usepackage{todonotes}

\newcommand{\B}{\mathbb{B}}
\newcommand{\R}{\mathbb{R}}

\newcommand{\apart}{\not\sim_{\ixT}}

\newcommand\MW{\textbf{[MW]}}
\newcommand\mc{\textbf{[m$^3$]}}
\newcommand\mcps{\textbf{[m$^3$/s]}}
\newcommand\meter{\textbf{[m]}}
\newcommand\doll{\textbf{[\$]}}
\newcommand\seco{\textbf{[s]}}
\newcommand\dpmw{\textbf{[\$$/$MW]}}
\newcommand\MWtMW{\textbf{[MW $\rightarrow$ MW]}}
\newcommand\mcpstm{\textbf{[m$^3$/s $\rightarrow$ m]}}
\newcommand\MWMWtMW{\textbf{[MW $\times$ MW $\rightarrow$ MW]}}

\newcommand\paramNumMaxGHpRsT[3]{\overline{\nu}_{#1,#2,#3}}
\newcommand\paramNumMinGHpRsT[3]{\underline{\nu}_{#1,#2,#3}}
\newcommand\paramPMaxGRsT[3]{\overline{\pi}_{#1,#2,#3}}
\newcommand\paramPMinGRsT[3]{\underline{\pi}_{#1,#2,#3}}
\newcommand\paramPMaxHpRsT[3]{\overline{\pi}_{#1,#2,#3}}
\newcommand\paramPMinHpRsT[3]{\underline{\pi}_{#1,#2,#3}}

\newcommand\paramRAS[2]{\rho_{#1,#2}}

\newcommand\paramPGCfgYDh[4]{\pi_{#1,#2}^{#3,#4}}

\newcommand\paramPZT[2]{\pi_{#1,#2}}
\newcommand\paramLReDh[2]{\mathit{\lambda}_{#1,#2}}
\newcommand\paramLamRiTTp[3]{\mathit{\rho}_{#1,#2,#3}}
\newcommand\paramPSgCfgYDh[4]{\pi_{#1,#2}^{#3,#4}}
\newcommand\paramUpperStabSgCfgYDh[4]{\upsilon^{\mathrm{trans}}_{#1,#2,#3,#4}}
\newcommand\paramSfcDown{\tau^\downarrow_{\ixSfc}}
\newcommand\paramSfcUp{\tau^\uparrow_{\ixSfc}}
\newcommand\paramSfcTot{\tau^\updownarrow_{\ixSfc}}
\newcommand\paramSfcDownSgCfgYDh[4]{\pi^{\downarrow,#3,#4}_{#1,#2}}
\newcommand\paramSfcUpSgCfgYDh[4]{\pi^{\uparrow,#3,#4}_{#1,#2}}
\newcommand\paramFGCfgYDh[4]{\phi_{#1,#2}^{#3,#4}}
\newcommand\paramVInitRe[1]{\nu_{#1}}
\newcommand\paramDurationT[1]{\delta_{#1}}
\newcommand\paramStabAbsSz[1]{\tau^{\mathrm{abs}}_{#1}}
\newcommand\paramStabRateSz[1]{\tau^{\mathrm{rate}}_{#1}}

\newcommand\bigMWorst{M^{\mathrm{worst}}}

\newcommand\costManCfgT[2]{\gamma^{\mathrm{commit}}_{#1,#2}}
\newcommand\costSetGT[2]{\gamma^{\mathrm{setpoint}}_{#1,#2}}

\newcommand\fctLossLiT[3]{\mathit{\Theta}_{#1,#2}(#3)}
\newcommand\fctLossMtdcZZpT[4]{\mathit{\Theta}_{\ixMtdc,#1,#2,#3}(#4)}
\newcommand\fctVLRe[2]{\mathit{\Gamma}_{#1}(#2)}

\newcommand\fctUpperNorthFcpl[1]{\mathit{\Upsilon}^{\mathrm{north}}(#1)}

\newcommand\fctPfcLimitT[2]{\mathit{\Lambda}^{\mathrm{pfc}}_{#1}(#2)}

\newcommand\ixChp{\mathit{ch}}
\newcommand\ixDh{\mathit{dh}}
\newcommand\ixCfg{\mathit{k}}
\newcommand\ixMtdc{\mathit{mtdc}}
\newcommand\ixFcpl{\mathit{fcpl}}
\newcommand\ixG{\mathit{g}}
\newcommand\ixGd{\mathit{g'}}

\newcommand\ixGp{\mathit{gp}}

\newcommand\ixHp{\mathit{h}}
\newcommand\ixIc{\mathit{ic}}
\newcommand\ixIr{\mathit{ir}}
\newcommand\ixLi{\mathit{li}}

\newcommand\ixRe{\mathit{re}}
\newcommand\ixRi{\mathit{ri}}
\newcommand\ixRorhp{\mathit{rh}}
\newcommand\ixRs{\mathit{r}}

\newcommand\ixSc{\mathit{sc}}
\newcommand\ixTc{\mathit{tc}}
\newcommand\ixSfc{\mathit{sfc}}
\newcommand\ixSg{\mathit{sg}}
\newcommand\ixSp{\mathit{sp}}
\newcommand\ixSz{\mathit{sz}}
\newcommand\ixT{\mathit{t}}
\newcommand\ixTz{\mathit{t_0}}
\newcommand\ixTp{\mathit{t'}}

\newcommand\ixY{\mathit{y}}
\newcommand\ixZ{\mathit{z}}
\newcommand\ixZp{\mathit{z'}}

\newcommand\ixRas{\mathit{ras}}

\newcommand\sE{\mathcal{E}}
\newcommand\sChp{\mathcal{CH}}
\newcommand\sCfg{\mathcal{K}}
\newcommand\sDh{\mathcal{DH}}
\newcommand\dhLow{\textrm{low}}
\newcommand\dhHi{\textrm{hi}}
\newcommand\sFcpl{\mathcal{FCPL}}
\newcommand\sG{\mathcal{G}}
\newcommand\sGNed{\mathcal{G}^{\mathrm{NED}}}

\newcommand\sGp{\mathcal{GP}}

\newcommand\sHp{\mathcal{H}}
\newcommand\sIc{\mathcal{Ic}}
\newcommand\sIr{\mathcal{Ir}}

\newcommand\sLi{\mathcal{Li}}

\newcommand\sRe{\mathcal{Re}}
\newcommand\sRi{\mathcal{Ri}}
\newcommand\sRs{\mathcal{R}}
\newcommand\sRorhp{\mathcal{RH}}

\newcommand\sTc{\mathcal{TC}}

\newcommand\sSp{\mathcal{Sp}}
\newcommand\sSz{\mathcal{SZ}}
\newcommand\sT{\mathcal{T}}
\newcommand\sTf{\mathcal{T}^+}

\newcommand\sY{\mathcal{Y}}
\newcommand\yMin{\mathit{min}}
\newcommand\yOpt{\mathit{opt}}
\newcommand\yMax{\mathit{max}}
\newcommand\yStab{\mathrm{stab}}
\newcommand\sYStab{\mathcal{Y}^{\yStab}}
\newcommand\sZ{\mathcal{Z}}
\newcommand\sZp{\mathcal{Z}'}
\newcommand\sRas{\mathcal{RAS}}

\newcommand\sCfgG[1]{\mathcal{K}_{#1}}
\newcommand\sCfgHp[1]{\mathcal{K}_{#1}}
\newcommand\sCfgSg[1]{\mathcal{K}_{#1}}
\newcommand\sChpZ[1]{\mathcal{CH}_{#1}}

\newcommand\sGHp[1]{\mathcal{G}_{#1}}

\newcommand\sTcSg[1]{\mathcal{SG}_{#1}}
\newcommand\sTcG[1]{\mathcal{G}_{#1}}
\newcommand\sFcplSg[1]{\mathcal{SG}_{#1}}
\newcommand\sGpZ[1]{\mathcal{GP}_{#1}}
\newcommand\mHpG[1]{\mathit{h}_{#1}}
\newcommand\sHpRe[1]{\mathcal{HP}_{#1}}
\newcommand\sHpRi[1]{\mathcal{HP}_{#1}}

\newcommand\sHpSz[1]{\mathcal{HP}_{#1}}
\newcommand\sIcZ[1]{\mathcal{Ic}_{#1}}
\newcommand\sIrZ[1]{\mathcal{Ir}_{#1}}
\newcommand\sInLiZ[1]{\mathcal{Li}^-_{#1}}
\newcommand\sOutLiZ[1]{\mathcal{Li}^+_{#1}}

\newcommand\mReSg[1]{\mathit{re}_{#1}}
\newcommand\sRiRe[1]{\mathcal{Ri}_{#1}}
\newcommand\sRorhpZ[1]{\mathcal{RH}_{#1}}
\newcommand\sSgT[1]{\mathcal{SG}_{#1}}
\newcommand\sSgHpT[2]{\mathcal{SG}_{#1,#2}}
\newcommand\mSgGT[2]{\mathit{sg}_{#1,#2}}
\newcommand\sSpRe[1]{\mathcal{SP}_{#1}}
\newcommand\sSpRi[1]{\mathcal{SP}_{#1}}
\newcommand\mTFcpl[1]{\mathit{t}_{#1}}

\newcommand\mTSg[1]{\mathit{t}_{#1}}
\newcommand\sFcplRT[2]{\mathcal{FCPL}_{#1,#2}}
\newcommand\sGFcpl[1]{\mathcal{G}_{#1}}

\newcommand\var[1]{\mathbf{#1}}
\newcommand\varActiveGRsT[3]{\var{A_{#1,#2,#3}}}
\newcommand\varPChpRsT[3]{\var{P_{#1,#2,#3}}}
\newcommand\varPInMtdcZRsT[3]{\var{P^-_{\ixMtdc,#1,#2,#3}}}
\newcommand\varPOutMtdcZRsT[3]{\var{P^+_{\ixMtdc,#1,#2,#3}}}
\newcommand\varPGRsT[3]{\var{P_{#1,#2,#3}}}
\newcommand\varPGpRsT[3]{\var{P_{#1,#2,#3}}}
\newcommand\varPIcRsT[3]{\var{P_{#1,#2,#3}}}
\newcommand\varPIrRsT[3]{\var{P_{#1,#2,#3}}}
\newcommand\varPInLiRsT[3]{\var{P^-_{#1,#2,#3}}}
\newcommand\varPOutLiRsT[3]{\var{P^+_{#1,#2,#3}}}

\newcommand\varPRorhpRsT[3]{\var{P_{#1,#2,#3}}}

\newcommand\varPPMRsT[2]{\var{P^{+,-}_{#1,#2}}}
\newcommand\varPWorstRT[2]{\var{P_{#1,#2}^{\mathrm{worst}}}}
\newcommand\varPSndWorstRT[2]{\var{P_{#1,#2}^{\mathrm{2nd}}}}
\newcommand\varPResRT[2]{\var{P_{#1,#2}^{\mathrm{res}}}}
\newcommand\varWorstFcplRT[3]{\var{b_{#1,#2,#3}^{\mathrm{worst}}}}
\newcommand\varSndWorstFcplRT[3]{\var{b_{#1,#2,#3}^{\mathrm{2nd}}}}
\newcommand\varbRAS[2]{\var{b_{#1,#2}}}

\newcommand\varACfgT[2]{\var{A_{#1,#2}}}

\newcommand\varAGT[2]{\var{A_{#1,#2}}}
\newcommand\varPGT[2]{\var{P_{#1,#2}}}
\newcommand\varFHpT[2]{\var{F_{#1,#2}}}
\newcommand\varPHpT[2]{\var{P_{#1,#2}}}

\newcommand\varLReT[2]{\var{L_{#1,#2}}}
\newcommand\varVReT[2]{\var{V_{#1,#2}}}
\newcommand\varVInRiT[2]{\var{V^-_{#1,#2}}}
\newcommand\varVOutRiT[2]{\var{V^+_{#1,#2}}}
\newcommand\varHSg[1]{\var{H_{#1}}}
\newcommand\varPSgY[2]{\var{P_{#1,#2}}}
\newcommand\varPTransSg[1]{\var{P^{\mathrm{trans}}_{#1}}}
\newcommand\varPSg[1]{\var{P_{#1}}}

\newcommand\varSfcDownHpT[2]{\var{P^\downarrow_{#1,#2}}}
\newcommand\varSfcUpHpT[2]{\var{P^\uparrow_{#1,#2}}}
\newcommand\varSfcDownSg[1]{\var{P^\downarrow_{#1}}}

\newcommand\varSfcUpSg[1]{\var{P^\uparrow_{#1}}}
\newcommand\varWSgCfgYDh[4]{\var{W_{#1,#2}^{#3,#4}}}
\newcommand\varVSpT[2]{\var{V_{#1,#2}}}

\newcommand\varPTransT[1]{\var{P^{\mathrm{trans}}_{#1}}}
\newcommand\varPWorstT[1]{\var{P_{#1}^{\mathrm{worst}}}}
\newcommand\varPPfcSg[1]{\var{P_{#1}^{\mathrm{pfc}}}}
\newcommand\exprUpperLimitLi[1]{\bm{\Upsilon}_{#1}}

\newcommand\exprLowerLimitInMtdcZRsT[3]{\bm{\Lambda}^-_{\ixMtdc,#1,#2,#3}}
\newcommand\exprUpperLimitInMtdcZRsT[3]{\bm{\Upsilon}^-_{\ixMtdc,#1,#2,#3}}
\newcommand\exprLowerLimitOutMtdcZRsT[3]{\bm{\Lambda}^+_{\ixMtdc,#1,#2,#3}}
\newcommand\exprUpperLimitOutMtdcZRsT[3]{\bm{\Upsilon}^+_{\ixMtdc,#1,#2,#3}}
\newcommand\exprLowerLimitInLiRsT[3]{\bm{\Lambda}^-_{#1,#2,#3}}
\newcommand\exprUpperLimitInLiRsT[3]{\bm{\Upsilon}^-_{#1,#2,#3}}
\newcommand\exprLowerLimitOutLiRsT[3]{\bm{\Lambda}^+_{#1,#2,#3}}
\newcommand\exprUpperLimitOutLiRsT[3]{\bm{\Upsilon}^+_{#1,#2,#3}}

\newcommand\exprLowerLimitMtdcT[1]{\bm{\Lambda}^-_{\ixMtdc,#1}}

\newcommand\exprUpperLimitSc[1]{\bm{\Upsilon}_{#1}}
\newcommand\exprUpperLimitTc[1]{\bm{\Upsilon}_{#1}}

\newcommand\paramcommitOrderg[1]{co_{#1}}

\newcommand\algforcedgroups{G^{\mathrm{fo}}}
\newcommand\algunavailgroups{G^{\mathrm{u}}}
\newcommand\algconfiggroups{K}
\newcommand\algpotentialgroups{G^{\mathrm{c}}}
\newcommand\algtoaddgroups{G}

\newcommand{\newtext}[1]{#1}

\makeatletter
\def\maketag@@@#1{\hbox{\m@th\normalfont\normalsize#1}}
\makeatother

\begin{document}

\title{Generation planning and operation under power stability constraints: A Hydro-Quebec use case}

\author{Alexandre Besner,
        Alexandre Blondin Mass\'e,
        Abderrahman Bani,
        Mouad Morabit,\\
        François Berthaut,
        Luc Charest,
        David Ialongo,
        Yves Mbeutcha,
        Simon Couture-Gagnon, and
        Julien Fournier
\thanks{A. Besner, A. Blondin Mass\'e, A. Bani, and M. Morabit
are with the Hydro-Quebec Research Institute (IREQ), Varennes, QC J3X 1S1,
Canada (e-mail:besner.alexandre@hydroquebec.com).}
\thanks{F. Berthaut, L. Charest, D. Ialongo, Y. A. Mbeutcha, S. Couture Gagnon and J. Fournier are with the Hydro-Quebec Operation and Planning Department, Montreal, QC H3A 3S7, Canada.}}



\maketitle

\begin{abstract}
  Hydro-Quebec (HQ) is a vertically integrated utility that produces, transmits, and distributes most of the electricity in the province of Quebec.
  The power grid it operates has a particular architecture created by large hydroelectric dams located far north and the extensive 735kV transmission grid that allows the generated power to reach the majority of the load located thousands of kilometers away in the southern region of Quebec.
  The specificity of the grid has led HQ to develop monitoring tools responsible for generating so-called stability limits.
  Those stability limits take into account several nonlinear phenomena such as angular stability, frequency stability, or voltage stability.
  Since generation planning and operation tools rely mostly on mixed integer linear programming formulation, HQ had to adapt its tools to integrate stability limits into them.
  This paper presents the challenges it faced, especially considering its reserve monitoring tool and unit commitment tool.
\end{abstract}


\section{Introduction}

Generation planning and operation are generally associated with canonical problems, such as the unit commitment (UC) problem and the optimal power flow (OPF) problem.
The UC can be formulated as a mixed integer and linear programming problem (MILP)~\cite{Anjos2017}.
To achieve a MILP formulation, the power flow constraints are often relaxed and validated \textit{a posteriori} using simulation.
For instance, Bender cuts can be generated if the simulation detects power flow violation in normal or post-contingency operation.
Similar security-constrained UC (SCUC) algorithms are used by system operators like ERCOT~\cite{Hui2013} or NYISO~\cite{NYISO2020} to balance their grid.
Unfortunately, this SCUC formulation cannot be adapted to Hydro-Quebec's grid, since it is primarily constrained by power system stability issues instead of line thermal ratings.
The computation power required to simulate power system stability phenomena is too intensive to generate cuts on demand for a MILP formulation~\cite{Huang2012}.
Moreover, the SCUC formulation is normally defined for thermal generation models.
To represent a grid whose generators are mainly hydroelectric, specific models are needed.
Hydro unit commitment (HUC) requires detailed modeling of the rivers network, reservoirs size and generator characteristics based on water level~\cite{taktak2017}.
Generator characteristics are often nonlinear and rely on linear approximation to be included in a MILP framework.
Some linearization approaches have been proposed~\cite{borghetti2008}.
While interesting, they do not address the complexity of cascading hydro plants. 
This paper describes how Hydro-Quebec models hydroelectric plants to account for both dynamic characteristics of generators and complex cascading river models.

Kundur et al.\ list of different power stability phenomena ingrouped in three groups: rotor angle stability, frequency stability, and voltage stability~\cite{Kundur2004}.
The latter two can be further divided into subcategories according to short-term or long-term phenomena.
Hydro-Quebec's grid is concerned with all three types of stability.
However, when looking at the literature about transient stability constrained unit commitment (TSCUC), only one category is usually taken into consideration.
For example, Ahmadi and Ghasemi consider a frequency stability-constrained version of the problem~\cite{Ahmadi2014}.
They transform the nonlinear frequency constraints into min-max constraints and subobjectives, then solve the problem using a MILP solver.
However, linearization other stability phenomena is not covered. 
Jiang, Zhou and Zhang address the rotor angle stability issue by proposing a TSCUC variant of the problem~\cite{Jiang2013}, using a parallelized augmented Lagrangian method to decompose the problem into multiple transient stability-constrained optimal power flows (TSCOPF).
Nevertheless, they report long computation timesas it takes several hours to solve a TSCUC for a 300-bus system.
This computation time is unacceptable in a real-time operation context.
Furthermore, their algorithm only considers one contingency.

Instead of identifying specific contingencies, another approach consists of establishing a long list of contingencies, since many factors can influence which one is the worst~\cite{Huang2012}.
For instance, in~\cite{Xu2015}, a decomposition framework to solve the same TSCUC with a rotor angle constraint is proposed as a way to consider multiple contingencies.
A disadvantage of this approach is that it relies on a transient stability assessment (TSA) to generate a cut for the master problem, and, as it was mentioned earlier, TSA is generally not fast enough to solve a problem in the context of real-time operation.
Finally, a mixed integer second-order cone programming formulation was suggested in~\cite{Chu2023}.
This approach considers voltage stability and frequency stability constraints.
However, the voltage stability constraints are mainly for inverter based resources (IBR) and are modeled as continuous constraints.
It does not take into account long-term voltage stability issues caused by transformer tap-change. 

Being part of the Northeast Power Coordination Council (NPCC), another important responsibility of Hydro-Quebec as a Reliability Coordinator (RC) is reserves monitoring.
Most of the NPCC members rely on zonal reserves threshold with no transmission constraints inside the zone~\cite{NYISO2023,ISONE2013,IESO2011}.
More precisely, they monitor their reserves by co-optimizing reserves and energy in their UC since the same resources often provide both services.
This strategy cannot be implemented by Hydro-Quebec, given that power is mainly generated far north while most of the load is consumed in its southern part.
Moreover, the activation of reserve resources can impact stability transfer limits on Hydro-Quebec, so it had to come up with an alternate approach to reserve monitoring.
For this purpose, Hydro-Quebec developed a novel transient stability constrained reserve monitoring (TSCRM) tool. 

This paper focuses on the unique approach used by Hydro-Quebec to include power stability constraints in the planning and operation of generation.
First, it describes how stability limits are calculated, prepared for the operators, and transformed into MILP constraints.
It then details the mathematical model behind Hydro-Quebec's TSCRM software, called ALFRED, and how it is used for operation tasks.
Finally, the mathematical model used to address the hydraulic TSCUC (HTSCUC) problem, implemented into the RALPH software, is introduced.
Since RALPH is still being actively under development, the paper also addresses present and future challenges posed by the HTSCUC problems.  

\section{Hydro-Quebec stability limits}

This section describes how Hydro-Quebec handles its stability limits.
First, Subsection~\ref{ss:grid-overview} summarizes the main particularity of Hydro-Quebec's power grid.
Next, Subsection~\ref{ss:simulations-required} provides more details about the LIMSEL tool and how it is used to estimate those stability limits.
Finally, Subsection~\ref{ss:limit-transformation} explains the process of transforming the stability limits into constraints in the MILP framework.

\subsection{Grid overview}\label{ss:grid-overview}


\begin{figure}
  \centering
  \includegraphics[width=8cm]{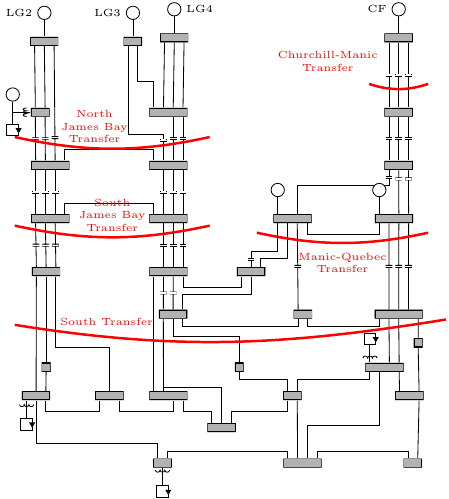}
  \caption{Hydro-Quebec’s main transmission network}\label{fig:HQ_grid}
\end{figure}

Figure~\ref{fig:HQ_grid} depicts the main components of HQ's transmission network, which includes radial 735kV transmission lines and a multi-terminal +/- 450kV DC line.
Many static Var compensators (SVC) and synchronous compensators (SC) are connected to HQ's grid.
HQ's network is not synchronized to the Eastern interconnection. Electrical connection with neighboring utilities is done through 6 back-to-back DC ties.
As a consequence, it does not have a strong inertia and it has higher frequency sensitivity,

The total installed capacity is about 48,000 MW including Churchill Falls (5,428 MW), 4,000 MW wind generation, and other independent power producers.
Around 90\% of the total capacity is hydroelectric and located north.
The principal load centers are located more than 1,000 km away in the south.
The historical peak demand is approximately 42,800 MW\@.

\subsection{Simulations required}\label{ss:simulations-required}

\begin{figure}[t]
  \centering
  \includegraphics[width=.93\linewidth]{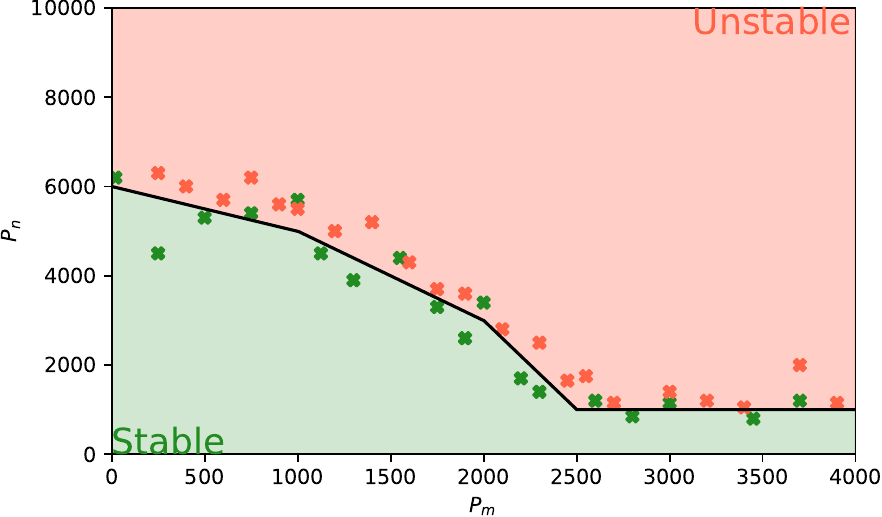}
  \caption{Stability limit of power interface $n$ for transit of power interface $m$}\label{fig:stability_limit}
\end{figure}

The simulations required to obtain the stability limits are detailed in~\cite{Huang2012}.
This process has remained similar since then.
Hydro-Quebec's grid needs various simulations to capture different stability issues like angular, frequency, voltage, and oscillations.
It is still hard to do contingency selection and ranking, as for different topologies, a power transfer interface can be constrained by different stability issues.
In addition, an aging grid has led to a significant rise in outages, either planned or not. It became necessary to increase the number of topologies covered and optimize the stability limits even more.
To accomplish this goal, OPAR, the PSS/E simulations management software used in \cite{Huang2012}, was replaced by DAR\@, a new in-house Python simulations management tool that allows the full customization and automation of \newtext{PSS/E} simulations preparation, parallel execution, key results storage, analysis, synthesis and output formatting.
It is a significant improvement in comparison with OPAR which could handle only a few simulations at a time, resulting in major stability limits studies (e.g. a power transfer interface) based only on a few thousands simulations results.
Also, tedious manual operations were needed before transferring the results to the operation engineers.
In contrast, DAR enables to perform and analyze unprecedented simulations volume. \newtext{Figure \ref{fig:stability_limit} presents an example of a fictive study conducted by DAR to identify the impact of the transit of power interface $m$ on the stability limit of the power interface $n$. Each cross on this figure represents a simulation scenario, composed of different elements.

Here is a non exhaustive list of parameters defining a scenario:
(1) \emph{Contingency}. Since multiple lines ar part of a power interface and HQ does not perform contingency ranking, it has to account for multiple contingencies.
(2) \emph{Study types}. PSS/E is used for all simulations but different parameters are needed based on the stability issues studied. For example, voltage stability studies need longer runtime to detect slow tap-changer actions. Meanwhile, angular stability can be verified significantly faster and need shorter simulations.
(3) \emph{Generation configuration}. When dealing with angular stability, generation configuration has a big impact on short-circuit level. Multiple configurations are tested to ensure the reliability of the stability limits.
(4) \emph{Load level}. Demand impacts the behavior of the grid especially for voltage collapse where MVAR consumption by the lines are influenced by load levels.
(5) \emph{Equipment outages}. Outages have a big impact on stability limits. They can affect the topology and the power flow or the transient response of the grid. It is essential to consider multiple combinations of outage while building simulation scenarios. 

After building the scenarios using DAR, the planning engineer will run the simulations and determine the stability limits based on the results. In the fictive example of Figure~\ref{fig:stability_limit}, the stability limit is represented by the black line using the input of the planning engineer. This process aims to be conservative and includes security margins resulting in some stable simulations not being included in the stable zone. The approach that systematically extracts stability limits based on simulation results is presented in~\cite{Valette2009}}

\begin{figure}[!t]
  \centering
  \includegraphics[width=.9\linewidth]{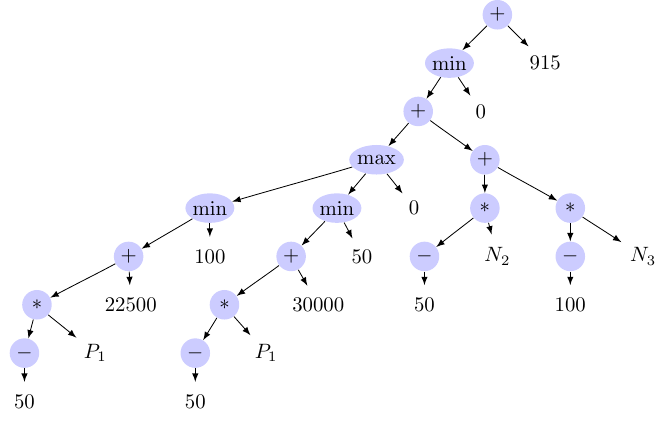}
  \caption{The tree of a limit expression}\label{fig:tree-limit}
\end{figure}

\newtext{In~\cite{Valette2009}, stability limits were formatted as a lookup table for quick real-time monitoring computations.
In this format, it was impossible to use the stability limits as input for the different MILP problems used in the operation planning horizon. With recent improvements, the lookup tables can now be represented as min-max expressions. Figure~\ref{fig:stability_limit} results can be converted into a min-max expression, such as the one presented in \eqref{eq:limite_figure}. It is important to notice that the results of DAR are generally not convex and can lead to inequations such as
\begin{equation}\label{eq:limite_figure}\begin{array}{rc@{\hspace{0pt}}l@{\hspace{0pt}}l}
  \exprUpperLimitLi{n} & \leq \max(1000, \min( & -P_m+6000, \\
                       &                       & -2P_m+7000, \\
                       &                       & -4P_m+11000))
\end{array}\end{equation}
}
\newtext{
  As another example, Figure~\ref{fig:tree-limit} shows the tree representation of the real min-max expression
\begin{equation}\label{eq:limit_tree}\begin{array}{rcl}
  \exprUpperLimitLi{n} & \leq & \min(\max(\min((-50)P_1 + 22~500, 100), \\
                       &      & \hphantom{\min(\max(}\mathrm{min}((-50)P_1 + 30~000, 50), 0) \\
                       &      & \hphantom{\min(}+ ((-50)N_2 + (-100)N_3), 0) + 915
\end{array}\end{equation}
produced by post-processing using DAR simulation results}, establishing a relation between the production of three distinct hydro plants $h_1$, $h_2$ and $h_3$,
where $P_1$ denotes the total production of hydro plant $h_1$, $N_2$ the number of generators of hydro plant $h_2$ that cannot contribute to the stability, 
and $N_3$ the number of generators of hydro plant $h_3$ that cannot participate in the stability.
Note that the tree of Figure~\ref{fig:tree-limit} is among the smallest expressions generated for the MILP problem: 
most of the remaining expressions are too large to be written or drawn in this paper.
A unique tree is generated for every stability limit and every topology studied.
As an example, the South transfer can have a tree with a depth of 49 and over 4900 nodes.


\subsection{Transforming limit expressions into linear constraints}
\label{ss:limit-transformation}

As described in Subsection~\ref{ss:simulations-required}, limit expressions are composed of piecewise linear nested min-max functions.
\newtext{
Many textbooks show that min-max constraints can be integrated in the MILP framework, to the cost of introducing additional binary variables and so-called big-M values, as well as modifying the objective function \cite{vielma2015mixed}.
However, if not carefully handled, these transformations of the initial problem can lead to \emph{ill-conditioned problem}, i.e. problems in which the magnitudes of the constant terms are too disparate, leading to numeric instability \cite{vielma2015mixed}.
}
Hence, although the use of the $\min$ and $\max$ functions within constraints is supported by some MILP solvers such as CPLEX, simple experimentation exploiting knowledge of the problem showed that a linearization done in a preprocessing step improves global performance.

Limit expressions are initially extracted and generated as raw strings.
The strings are then parsed into an abstract syntax tree (AST), so that some transformations and simplifications can be performed more easily.
During this normalizing process, each function $f(\sE) \in{} \{+,-,\cdot,/,\min,\max\}$ is mapped to internal tree node, while its subexpression arguments $e_i\in{}\sE$ are mapped to subtrees (Figure~\ref{fig:tree-limit}).
\newtext{
Then transformative simplifications can be applied on the tree to reduce its size~\cite{cuninghame2012minimax}, based on known arithmetical identities.
These transformations include (1) reordering additions and multiplications in virtue of their commutativity, (2) propagating the arithmetic inverse to the leaf nodes, (3) applying the identities $-\min(\sE)\equiv{}\max(-\sE)$ and $-\max(\sE)\equiv{}\min(-\sE)$, (4) merging constant values, (5) elimininating redundant subtrees and (5) factoring out common subtrees.
}
Next, a bound analysis is performed so that some branches of the tree can be pruned.
As an illustration, with the subexpression $\min(100,-50P_1+22500)$ in Figure~\ref{fig:tree-limit}, if it is known that the domain range of $P_1 \in [0,400]$ (the exact value of the variable being unknown), then the image range of expressions $100$ and $-50P_1+22500$ are respectively bounded by $[100,100]$ and $[2500,22500]$.
Since the first image range dominates the second (there can be no value $v\in[2500,22500]$ for which it would be smaller than a value within the first range), then the second subexpression can be safely eliminated.
Moreover, once a branch has been pruned, further simplification can be triggered recursively.
Experiments show that those preprocessing steps can significantly reduce the complexity of the tree met in real-life datasets.
Another useful simplification is obtained by identifying identical subtrees, which can be isolated and extracted into other constraints simply by introducing intermediate variables.
Continuing on the example of Figure~\ref{fig:tree-limit}, since the subexpression $-50P_1$ is repeated twice, each of its occurrences could be replaced with a new variable and a new constraint.

Once the main simplification and bounds analysis have been applied, the linearisation of $\min$/$\max$ are done as
follows:

\footnotesize
\begin{equation}\label{eq:maxsansbin}
  x\ge{}\max(\sE) \Rightarrow x\ge{}e_i, \forall e_i \in \sE
\end{equation}
\begin{equation}\label{eq:minsansbin}
  x\le{}\min(\sE) \Rightarrow x\le{}e_i, \forall e_i \in \sE
\end{equation}
\begin{equation}\label{eq:maxavecbin}
  x\le{}\max(\sE)\Rightarrow
  \begin{cases}
  d\le{}e_i+M-Mb_{e_i}, \forall e_i \in \sE \\
  x=d,\\
  \sum_{e_i \in \sE}{b_{e_i}}=1
  \end{cases}
\end{equation}
\begin{equation}\label{eq:minavecbin}
  x\ge{}\min(\sE) \Rightarrow
  \begin{cases}
  d\ge{}e_i-M+Mb_{e_i}, \forall e_i \in \sE \\
  x=d,\\
  \sum_{e_i\in \sE}{b_{e_i}}=1,
  \end{cases}
\end{equation}
\normalsize

where $M$ is a well-chosen constant (usually called \emph{big-M} in MILP), $d$ is a continuous variable and $b_{e_i}\in\B$ are binary variables.
Although binary variables put additional stress on the solver, several inequalities end up with only a few of them.


\newtext{
\section{Operation planning at HQ}
Operation planning can be described as all the verifications done prior to real-time operation to ensure smooth operation of the grid. At HQ, the operation window is typically split in four time frames.
1) \emph{1 year}.
Transmission and generation owners need to plan their major outages in advance.
During this horizon, operation planning must be done to ensure sufficient capacity and realizability of the requested outage.
2) \emph{28 days}.
This is the period where detailed outage studies are initiated. Transmission and generation owners have to provide detailed outage procedure. Detailed studies are conducted to ensure that the outages planned during the 1 year period are still feasible considering current grid condition and unplanned outages.
3) \emph{Day ahead}.
The day ahead period is used to validate if all the outages schedules and generation plan are still possible based on the latest and more precise forecasts.
It is also during this horizon that commercial transactions on interconnections are starting to be negotiated.
4) \emph{Real-time}.
Some studies are done during real-time operation to respond to unplanned events and correct the situation in the following minutes or hours depending on the severity of the event.
\begin{figure}
    \centering
    \includegraphics[scale=0.48]{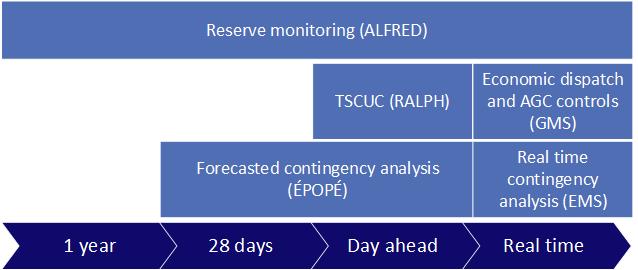}
    \caption{Operation planning at HQ overview}
    \label{fig:OperationPlanning}
\end{figure}
Figure \ref{fig:OperationPlanning} presents the horizons in which each tool is used.
ALFRED (see section \ref{sec:alfred}) is used in every horizon.
Using statistical data, ALFRED is a great tool to ensure sufficient capacity and reserve when planning long term outages.
The closer it becomes to real-time operation, the more it uses forecast data instead to ensure sufficient capacity and reserve for generation.
RALPH (see section \ref{sec:ralph}) is used to run the TSCUC problem during day-ahead planning to ensure that the generation will respect electric and hydraulic operational constraints.
ÉPOPÉ is a tool in development used to run contingency analysis solely based on forecast data instead of historical data.
It is mainly used in the day-ahead horizon, but also in the 28 days horizon for outage approval.
This paper does not address ÉPOPÉ, but further details can be found in \cite{delavari2024}.

}
\section{ALFRED\@: Monitoring and restoring reserves}\label{sec:alfred}

ALFRED is a tool used to monitor reserve levels and determine remedial actions to restore reserve levels when deficits are detected.
To do so, it solves one optimization problem for each reserve.
This section details the general model of the power adequacy problem used to determine if sufficient reserve resources are available.

All model components discussed in the section are represented as sets, as described in Table~\ref{tab:alfred-sets}, where a specific component is identified by an index.
For instance, the set of all hydro generators is denoted by $\sG$ and a specific hydro generator is represented by the index $\ixG$.
The relations between model components are also represented either by sets with indices or by an index with indices (Table~\ref{tab:alfred-sets}).
The parameters, \textit{i.e.} constant values for the optimization problem, are listed in Table~\ref{tab:alfred-parameters}, while the description of all decision variables is available in Table~\ref{tab:alfred-variables}.
To distinguish between parameters and variables, Greek letters are used for the former and bold capital letters for the latter.
Finally, functions (linear or piecewise linear) are represented as capital Greek letters (Table~\ref{tab:alfred-parameters}), and variable expressions (linear or piecewise) are identified with bold capital Greek letters (Table~\ref{tab:alfred-variables}).

\begin{table}[t]
  \caption{Indices, sets and relations}\label{tab:alfred-sets}
  \centering
  \scriptsize
  \begin{tabular}{>{\centering\arraybackslash}r@{\hspace{1mm}}c@{\hspace{1mm}}lm{5.4cm}}
    \toprule
    Index & & Set & Description \\
    \midrule
    $\ixFcpl$ & $\in$ & $\sFcpl$
      & Set of all first contingency production loss (FCPL) sets \\
    $\ixChp$ & $\in$ & $\sChp$
      & Set of all controllable hydro plants \\
    $\ixRorhp$ & $\in$ & $\sRorhp$
      & Set of all run-of-river hydro plants \\
    $\ixHp$ & $\in$ & $\sHp$
      & Set of all hydro plants ($\sHp = \sChp \cup \sRorhp$) \\
    $\ixG$ & $\in$ & $\sG$
      & Set of all hydro generators \\
    $\ixGp$ & $\in$ & $\sGp$
      & Set of all gas plants \\
    $\ixIc$ & $\in$ & $\sIc$
      & Set of all interconnectors \\
    $\ixIr$ & $\in$ & $\sIr$
      & Set of all interruptible \\
    $\ixLi$ & $\in$ & $\sLi$
      & Set of all links \\
    $\ixRas$ & $\in$ & $\sRas$
      & Set of all remedial action schemes \\
    $\ixRs$ & $\in$ & $\sRs$
      & Set of all reserves \\
    $\ixT$ & $\in$ & $\sT$
      & Set of all time steps (including initial time step $\ixTz$) \\
    $\ixT$ & $\in$ & $\sTf$
      & Set of all future time steps (excluding $\ixTz$) \\
    $\ixTc$ & $\in$ & $\sTc$
      & Set of all topology constraints \\
    $\ixZ$ & $\in$ & $\sZ$
      & Set of all zones \\
    $\ixZ$ & $\in$ & $\sZp$
      & Set of all zones excluding the south zone \\
    \bottomrule
  \end{tabular}
  \vskip\baselineskip
  \begin{tabular}{m{1.0cm}m{6cm}}
    \toprule
    Set & Description \\
    \midrule
    $\sFcplRT{\ixRs}{\ixT}$
      & Set of all FCPL active for reserve $\ixRs$ at time step $\ixT$ \\
    $\sGHp{\ixHp}$ \
      & Set of all hydro generators of hydro plant $\ixHp$ \\
    $\sChpZ{\ixZ}$
      & Set of all controllable hydro plants of zone $\ixZ$ \\
    $\sRorhpZ{\ixZ}$
      & Set of all run-of-river hydro plants of zone $\ixZ$ \\
    $\sGpZ{\ixZ}$
      & Set of all gas plants of zone $\ixZ$ \\
    $\sIcZ{\ixZ}$
      & Set of all interconnectors in zone $\ixZ$ \\
    $\sIrZ{\ixZ}$
      & Set of all interruptible loads of zone $\ixZ$ \\
    $\sInLiZ{\ixZ}$
      & Set of all incoming links of zone $\ixZ$ \\
    $\sOutLiZ{\ixZ}$
      & Set of all outgoing links of zone $\ixZ$ \\
    \bottomrule
  \end{tabular}
\end{table}

\begin{table}[t]
  \caption{Parameters and functions}\label{tab:alfred-parameters}
  \centering
  \scriptsize
  \begin{tabular}{m{1.8cm}m{6.0cm}}
    \toprule
    Parameter & Description \\
    \midrule
    $\paramPMinGRsT{\ixG}{\ixRs}{\ixT}$/
    $\paramPMaxGRsT{\ixG}{\ixRs}{\ixT}$
      & Minimum/maximum power produced by hydro generator $\ixG$ for reserve $\ixRs$
        during time step $\ixT$ \MW{} \\
    $\paramNumMinGHpRsT{\ixHp}{\ixRs}{\ixT}$/
    $\paramNumMaxGHpRsT{\ixHp}{\ixRs}{\ixT}$
      & Minimum/maximum number of active hydro generators in hydro plant $\ixHp$
        for reserve $\ixRs$ during time step $\ixT$ \MW{} \\
    $\paramPMinHpRsT{\ixHp}{\ixRs}{\ixT}$/
    $\paramPMaxHpRsT{\ixHp}{\ixRs}{\ixT}$
      & Minimum/maximum power produced by hydro plant $\ixHp$ for reserve $\ixRs$
        during time step $\ixT$ \MW{} \\
    $\paramPZT{\ixZ}{\ixT}$
      & Non dispatchable net power (bio-gas/private/wind planned generation
        and forecasted load) for zone $\ixZ$ during time step $\ixT$ \MW{} \\
    \midrule
    $\fctLossMtdcZZpT{\ixZ}{\ixZp}{\ixT}{\cdot}$
      & Linear function correcting a power transiting
        through the MTDC from zone $\ixZ$ into zone $\ixZp$
        at time step $\ixT$ by removing its transit loss \MWtMW{} \\
    \midrule
    $\fctLossLiT{\ixLi}{\ixT}{\cdot}$
      & Linear function correcting a power transiting
        through link $\ixLi$ at time step $\ixT$ by removing
        its loss \MWtMW{} \\
    \bottomrule
  \end{tabular}
\end{table}

\begin{table}[t]
  \caption{Decision variables and expressions}\label{tab:alfred-variables}
  \centering
  \scriptsize
  \begin{tabular}{m{1.9cm}m{5.9cm}}
    \toprule
    Variable & Description \\
    \midrule
    $\varWorstFcplRT{\ixFcpl}{\ixRs}{\ixT}$/
    $\varSndWorstFcplRT{\ixFcpl}{\ixRs}{\ixT}$
      & Indicator if $\ixFcpl$ is the worst/second worst FCPL for reserve
        $\ixRs$ during time step $\ixT$ \\
    $\varActiveGRsT{\ixG}{\ixRs}{\ixT}$
      & Indicator if hydro generator $\ixG$ is active for reserve $\ixRs$
        during time step $\ixT$ \\
    $\varbRAS{\ixRas}{\ixT}$
      & Indicator if remedial action scheme $\ixRas$ is applied at time step
        $\ixT$ \\
    \midrule
    $\varPPMRsT{\ixRs}{\ixT}$
      & The surplus/deficit of production in the south zone for reserve $\ixRs$
        during time step $\ixT$ \MW{} \\
    $\varPWorstRT{\ixRs}{\ixT}$/
    $\varPSndWorstRT{\ixRs}{\ixT}$
      & Production of the worst/second worst FCPL for reserve $\ixRs$ during
        time step $\ixT$ \MW{} \\
    $\varPResRT{\ixRs}{\ixT}$
      & Required power reserve for reserve $\ixRs$ during time step $\ixT$ \MW{}
        \\
    $\varPChpRsT{\ixChp}{\ixRs}{\ixT}$
      & Power produced by controllable hydro plant $\ixChp$ for reserve $\ixRs$
        during time step $\ixT$ \MW{} \\
    $\varPInMtdcZRsT{\ixZ}{\ixRs}{\ixT}$/
    $\varPOutMtdcZRsT{\ixZ}{\ixRs}{\ixT}$
      & Power transiting from zone $\ixZ$ into the MTDC/from the MTDC into
        zone for reserve $\ixRs$ during time step $\ixT$ \MW{} \\
    $\varPGRsT{\ixG}{\ixRs}{\ixT}$
      & Power produced by hydro generator $\ixG$ for reserve $\ixRs$ during time step
        $\ixT$ \MW{} \\
    $\varPGpRsT{\ixGp}{\ixRs}{\ixT}$
      & Power produced by gas plant $\ixGp$ for reserve $\ixRs$ during time step
        $\ixT$ \MW{} \\
    $\varPIcRsT{\ixIc}{\ixRs}{\ixT}$
      & Power coming from interconnector $\ixIc$ for reserve $\ixRs$ during time
        step $\ixT$ \MW{} \\
    $\varPIrRsT{\ixIr}{\ixRs}{\ixT}$
      & Power shedded by interruptible $\ixIr$ for reserve $\ixRs$ during time
        step $\ixT$ \MW{} \\
    $\varPInLiRsT{\ixLi}{\ixRs}{\ixT}$/
    $\varPOutLiRsT{\ixLi}{\ixRs}{\ixT}$
      & Power coming into/leaving link $\ixLi$ for reserve $\ixRs$ during time
        step $\ixT$ \MW{} \\
    $\varPRorhpRsT{\ixRorhp}{\ixRs}{\ixT}$
      & Power produced by run-of-river hydro plant $\ixRorhp$ for reserve $\ixRs$
        during time step $\ixT$ \MW{} \\
    \bottomrule
  \end{tabular}
  \vskip\baselineskip
  \begin{tabular}{m{1.4cm}m{6.4cm}}
    \toprule
    Expression & Description \\
    \midrule
    $\exprLowerLimitInMtdcZRsT{\ixZ}{\ixRs}{\ixT}$/
    $\exprUpperLimitInMtdcZRsT{\ixZ}{\ixRs}{\ixT}$
      & Piecewise linear expression representing the lower/upper
        limit of the power coming from zone $\ixZ$ into the MTDC
        for reserve $\ixRs$ at time step $\ixT$ \MW{} \\
    \midrule
    $\exprLowerLimitOutMtdcZRsT{\ixZ}{\ixRs}{\ixT}$/
    $\exprUpperLimitOutMtdcZRsT{\ixZ}{\ixRs}{\ixT}$
      & Piecewise linear expression representing the lower/upper
        limit of the power coming from the MTDC into zone $\ixZ$
        for reserve $\ixRs$ at time step $\ixT$ \MW{} \\
    \midrule
    $\exprLowerLimitInLiRsT{\ixLi}{\ixRs}{\ixT}$/
    $\exprUpperLimitInLiRsT{\ixLi}{\ixRs}{\ixT}$
      & Piecewise linear expression representing the lower/upper
        limit of power going into link $\ixLi$ for reserve $\ixRs$
        at time step $\ixT$ \MW{} \\
    \midrule
    $\exprLowerLimitOutLiRsT{\ixLi}{\ixRs}{\ixT}$/
    $\exprUpperLimitOutLiRsT{\ixLi}{\ixRs}{\ixT}$
      & Piecewise linear expression representing the lower/upper
        limit of power coming from link $\ixLi$ for reserve $\ixRs$
        at time step $\ixT$ \MW{} \\
    \midrule
    $\exprUpperLimitTc{\ixTc}$
      & Piecewise linear expression representing the upper limit of power
        produced by hydro generators in topology constraint $\ixTc$ \MW{} \\
    \bottomrule
  \end{tabular}
\end{table}

\subsection{Mathematical formulation}

The power adequacy problem is time and reserve independent.
It means that, if ALFRED has to monitor a set $R$ of reserves over a set $T$ of time steps, then $|R||T|$ power adequacy problems need to be solved.
If there are not enough resources to maintain a minimum reserve level, a corrective power adequacy problem is solved to determine actions to restore reserve levels. However, the corrective power adequacy problem is solved once for all reserves at a specific time step $\ixT$ since remedial actions are applied to all reserve types.

\subsubsection{Generation}

The power adequacy problem aims to maximize available generation at a specific time $t$.
Therefore, the generation model is simpler than a traditional unit commitment model.
Also, it neglects hydraulic constraints, which induce time dependencies, since it only monitors instantaneous power.
Those simplifications allow for the simple model expressed by Equations~\eqref{eqGenAlfred}-\eqref{eqGenCentraleAlfred}:

\footnotesize
\begin{equation}
  \label{eqGenAlfred}
  \begin{aligned}
    \paramPMinGRsT{\ixG}{\ixRs}{\ixT}\varActiveGRsT{\ixG}{\ixRs}{\ixT}
      \leq \varPGRsT{\ixG}{\ixRs}{\ixT}
      \leq \paramPMaxGRsT{\ixG}{\ixRs}{\ixT}\varActiveGRsT{\ixG}{\ixRs}{\ixT} \quad \forall \ixG\;\forall \ixRs\;\forall \ixT
  \end{aligned}
\end{equation}
\begin{equation}
  \label{eqNbGroupesAlfred}
  \begin{aligned}
    \paramNumMinGHpRsT{\ixHp}{\ixRs}{\ixT}
      \leq \sum_{\ixG \in \sGHp{\ixHp}} \varActiveGRsT{\ixG}{\ixRs}{\ixT}
      \leq \paramNumMaxGHpRsT{\ixHp}{\ixRs}{\ixT} \quad \forall \ixHp\;\forall \ixRs\;\forall \ixT
  \end{aligned}
\end{equation}
\begin{equation}
  \label{eqGenCentraleAlfred}
  \begin{aligned}
    \paramPMinHpRsT{\ixHp}{\ixRs}{\ixT}
      \leq \sum_{\ixG \in \sGHp{\ixHp}} \varPGRsT{\ixG}{\ixRs}{\ixT}
      \leq \paramPMaxHpRsT{\ixHp}{\ixRs}{\ixT} \quad \forall \ixHp\;\forall \ixRs\;\forall \ixT
  \end{aligned}
\end{equation}
\normalsize

Equation~\eqref{eqGenAlfred} constrains the power produced by a single hydro generator with respect to its state.
Equations~\eqref{eqNbGroupesAlfred} and~\eqref{eqGenCentraleAlfred} represent the operational constraints within each hydro plant.
In particular, a hydro plant can be limited in its number of hydro generators because of operational constraints or outages.
Similarly, hydro plants can be constrained in total power because of constraints on minimum or maximum water flows that can be turbinated.

\subsubsection{Transmission}

The following equations represent the model of the transmission grid: 

\footnotesize
\begin{align}
  & \begin{aligned}\label{eqPertesLienAlfred}
    \varPOutLiRsT{\ixLi}{\ixRs}{\ixT} & =
      \fctLossLiT{\ixLi}{\ixT}{\varPInLiRsT{\ixLi}{\ixRs}{\ixT}}
  & \forall \ixLi\;\forall \ixRs\; \forall \ixT
  \end{aligned} \\
  & \begin{aligned}\label{eqLimite1LienAlfred}
    \exprLowerLimitInLiRsT{\ixLi}{\ixRs}{\ixT} & \leq
    \varPInLiRsT{\ixLi}{\ixRs}{\ixT} \leq
    \exprUpperLimitInLiRsT{\ixLi}{\ixRs}{\ixT}
  & \forall \ixLi\;\forall \ixRs\;\forall \ixT
  \end{aligned} \\
  & \begin{aligned}\label{eqLimite2LienAlfred}
    \exprLowerLimitOutLiRsT{\ixLi}{\ixRs}{\ixT} & \leq
    \varPOutLiRsT{\ixLi}{\ixRs}{\ixT} \leq
    \exprUpperLimitOutLiRsT{\ixLi}{\ixRs}{\ixT}
  & \forall \ixLi\;\forall \ixRs\;\forall \ixT
  \end{aligned} \\
  & \begin{aligned}\label{eqPertesMtdcAlfred}
    \varPOutMtdcZRsT{\ixZ}{\ixRs}{\ixT} & =
      \fctLossMtdcZZpT{\ixZ}{\ixZp}{\ixT}{\varPInMtdcZRsT{\ixZp}{\ixRs}{\ixT}}
  & \forall \ixZ\;\forall \ixZp\;\forall \ixRs\;\forall \ixT \\
  \end{aligned} \\
  & \begin{aligned}\label{eqLimite1MtdcAlfred}
    \exprLowerLimitInMtdcZRsT{\ixZ}{\ixRs}{\ixT} & \leq
    \varPInMtdcZRsT{\ixZ}{\ixRs}{\ixT} \leq
    \exprUpperLimitInMtdcZRsT{\ixZ}{\ixRs}{\ixT}
  & \forall \ixZ\;\forall \ixRs\;\forall \ixT
  \end{aligned} \\
  & \begin{aligned}\label{eqLimite2MtdcAlfred}
    \exprLowerLimitOutMtdcZRsT{\ixZ}{\ixRs}{\ixT} & \leq
    \varPOutMtdcZRsT{\ixZ}{\ixRs}{\ixT} \leq
    \exprUpperLimitOutMtdcZRsT{\ixZ}{\ixRs}{\ixT}
  & \forall \ixZ\;\forall \ixRs\;\forall \ixT
  \end{aligned}
\end{align}
\normalsize

A linear approximation is used to represent the power traveling inside the transmission grid.
The zones are connected by links, i.e.\ aggregation of transmission lines between two zones.
Because of the particular architecture of Hydro-Quebec's grid (see Figure \ref{fig:HQ_grid}), the links are mostly unidirectional, which allows the identification of an upstream and a downstream extremity and neglects impedance.
Equation~\eqref{eqPertesLienAlfred} details the relationship between the upstream and downstream flow of a link. 
Links are also constrained by the stability limits described in Section~\ref{ss:limit-transformation}, which are expressed through Equations~\eqref{eqLimite1LienAlfred} and~\eqref{eqLimite2LienAlfred}.
Moreover, Hydro-Quebec operates a multi-terminal DC (MTDC) grid, whose specific operational constraints require more complex modeling.
To preserve the simplicity of the model in this article, the MTDC is represented as a special link, leading to Relations~\eqref{eqPertesMtdcAlfred}, \eqref{eqLimite1MtdcAlfred} and \eqref{eqLimite2MtdcAlfred}.

Equation~\eqref{eqEquilibreZone1} ensures that the power is balanced in each zone, except for the south zone.

\footnotesize
\begin{align}
    \label{eqEquilibreZone1}
    & \begin{aligned} 
      \paramPZT{\ixZ}{\ixT} =
      & \sum_{\ixChp \in \sChpZ{\ixZ}} \varPChpRsT{\ixChp}{\ixRs}{\ixT}
        + \sum_{\ixGp \in \sGpZ{\ixZ}} \varPGpRsT{\ixGp}{\ixRs}{\ixT}
        + \sum_{\ixIc \in \sIcZ{\ixZ}} \varPIcRsT{\ixIc}{\ixRs}{\ixT} \\
      & - \sum_{\ixIr \in \sIrZ{\ixZ}} \varPIrRsT{\ixIr}{\ixRs}{\ixT}
        + \sum_{\ixLi \in \sInLiZ{\ixZ}} \varPOutLiRsT{\ixLi}{\ixRs}{\ixT}
        - \sum_{\ixLi \in \sOutLiZ{\ixZ}} \varPInLiRsT{\ixLi}{\ixRs}{\ixT} \\
      & + \sum_{\ixRorhp \in \sRorhpZ{\ixZ}} \varPRorhpRsT{\ixRorhp}{\ixRs}{\ixT}
        + \varPOutMtdcZRsT{\ixZ}{\ixRs}{\ixT}
        - \varPInMtdcZRsT{\ixZ}{\ixRs}{\ixT} \\
      & \quad \forall \ixRs\;\forall \ixT\;\forall \ixZ \in \sZp \\
    \end{aligned}
\end{align}
\normalsize

Topology-based constraints (TC) are also monitored.
A topology processing tool is used to determine if a certain generation subset can create overload on the transformer or individual line.
All cases detected are used to build the TC set:

\footnotesize
\begin{align}
    \label{eqContrainteEnsembleAlfred}
    & \begin{aligned} 
       &\sum_{\ixG \in \sTcG{\ixTc}} \varPGRsT{\ixG}{\ixRs}{\ixT} \leq \exprUpperLimitTc{\ixTc} 
        &\forall \ixRs\;\forall \ixT\;\forall  \ixTc
      \end{aligned}  
\end{align}
\normalsize

\subsubsection{Reserves}

\newtext{To achieve monitoring reserve levels, reserve requirements are integrated directly into the model, each reserve being solved in its power adequacy problem. HQ has 9 reserves modeled in ALFRED that can be divided in 3 categories: (1) normal reserves, (2) stability reserves and (3) hydraulic availability reserves.
\emph{Normal reserves} are standard operation reserves defined by the NPCC's Directory 5~\cite{NPCC2019} like 10 minutes spinning reserve (10S), 10-minutes non-spinning reserve (10NS) and 30-minutes non-spinning reserve (30NS).
In addition, HQ adapted those reserves to specific used case.
The firm 30NS reserve (F30NS) is similar to the 30NS reserve but ignores non-firm transactions on the different tie lines, allowing HQ to assess its dependency to neighboring grids.
\emph{Stability reserve} ensures sufficient primary frequency control reserve (PFC) to cover for the worst first contingency production loss (FCPL) to ensure that the frequency deviation will be kept below the under frequency load shedding automaton threshold.
\emph{Hydraulic availability reserves} ensure that HQ has the capacity to deliver power in a sustainable manner.
It focuses more on operational and economics decision than reliability like the other two categories.
There are 4 reserves in this category: basic hydraulic availability (HA), optimal hydraulic availability (OptHA), where the maximal output is capped at the optimal yield of the generators (see Section~\ref{ss:hydraulic} for details on optimal yield of a generator), economic hydraulic availability (EcoHA) and low net demand hydraulic availability (LowHA).
LowHA is a new approach trying to capture emerging issues caused by low demand with high penetration of renewables. It is still a work in progress and is actively being worked on as the knowledge on those emerging issues increases.
In order to keep the model simple, only the reserves described by Directory 5 are detailed in this paper.}
This standard defines the operating reserves required as a function of the worst FCPL and the second worst one.
This translates into the problem as follows (where $\bigMWorst$ is some well-chosen big-M value):

\footnotesize
\begin{align}  
  &\begin{aligned} \label{eqWorst}
    \sum_{\ixFcpl\in \sFcplRT{\ixRs}{\ixT}} \varWorstFcplRT{\ixFcpl}{\ixRs}{\ixT} = 1 \quad \forall \ixRs \; \forall \ixT
  \end{aligned} \\
  &\begin{aligned} \label{eq2nd}
    \sum_{\ixFcpl\in \sFcplRT{\ixRs}{\ixT}} \varSndWorstFcplRT{\ixFcpl}{\ixRs}{\ixT} = 1 \quad \forall \ixRs \; \forall \ixT
  \end{aligned} \\
  &\begin{aligned} \label{AlfredWorstInf}
    \varPWorstRT{\ixRs}{\ixT} \geq \sum_{\ixG \in \sGFcpl{\ixFcpl}} \varPGRsT{\ixG}{\ixRs}{\ixT} \quad \forall \ixFcpl \; \forall \ixRs \; \forall \ixT
  \end{aligned} \\
  &\begin{aligned} \label{AlfredWorstSup}
    \varPWorstRT{\ixRs}{\ixT} \leq \sum_{\ixG \in \sGFcpl{\ixFcpl}} \varPGRsT{\ixG}{\ixRs}{\ixT} + (1-\varWorstFcplRT{\ixFcpl}{\ixRs}{\ixT}) \bigMWorst \quad \forall \ixFcpl \; \forall \ixRs \; \forall \ixT
  \end{aligned} \\
  &\begin{aligned} \label{Alfred2ndInf}
    \varPSndWorstRT{\ixRs}{\ixT} \geq \sum_{\ixG \in \sGFcpl{\ixFcpl}} \varPGRsT{\ixG}{\ixRs}{\ixT}-\bigMWorst \varWorstFcplRT{\ixFcpl}{\ixRs}{\ixT} \quad \forall \ixFcpl \; \forall \ixRs \; \forall \ixT
  \end{aligned} \\
  &\begin{aligned}\label{Alfred2ndSup} 
    \varPSndWorstRT{\ixRs}{\ixT} \leq \sum_{\ixG \in \sGFcpl{\ixFcpl}} \varPGRsT{\ixG}{\ixRs}{\ixT} + (1-\varSndWorstFcplRT{\ixFcpl}{\ixRs}{\ixT}) \bigMWorst \quad \forall \ixFcpl \; \forall \ixRs \; \forall \ixT
  \end{aligned} \\
  &\begin{aligned} \label{AlfredFCPLUnique} 
      \varWorstFcplRT{\ixFcpl}{\ixRs}{\ixT} + \varSndWorstFcplRT{\ixFcpl}{\ixRs}{\ixT} \leq 1 \quad \forall \ixFcpl \; \forall \ixRs \; \forall \ixT
  \end{aligned}\\
  &\begin{aligned} \label{Alfred10S} 
    \varPResRT{\ixRs=\mathrm{10S}}{\ixT} = 0.25\varPWorstRT{\ixRs=\mathrm{10S}}{\ixT}  \quad \forall \ixT
  \end{aligned}\\
  &\begin{aligned} \label{Alfred10NS} 
    \varPResRT{\ixRs=\mathrm{10NS}}{\ixT} = \varPWorstRT{\ixRs=\mathrm{10NS}}{\ixT} \quad \forall \ixT
  \end{aligned}\\
  &\begin{aligned} \label{Alfred30NS} 
    \varPResRT{\ixRs=\mathrm{30NS}}{\ixT} = \varPWorstRT{\ixRs=\mathrm{30NS}}{\ixT} + 0.5\varPSndWorstRT{\ixRs=\mathrm{30NS}}{\ixT}  \quad \forall \ixT
  \end{aligned}
\end{align}
\normalsize

Equations \eqref{eqWorst} and \eqref{eq2nd} guarantee that there is exactly one worst FCPL and exactly one second worst FCPL.
The FCPL set is created with the same topology processing tool as the TC set.
It identifies which hydro generator can be lost on a single contingency and builds the set dynamically.
Equations \eqref{AlfredWorstInf} and \eqref{AlfredWorstSup} determine the power generation value from the worst FCPL.
Equations \eqref{Alfred2ndInf} and \eqref{Alfred2ndSup} do the same, but for the second worst FCPL.
Equation \eqref{AlfredFCPLUnique} makes sure that two different contingencies are selected for the worst and second worst.
Equations \eqref{Alfred10S}-\eqref{Alfred30NS} determine the required reserve level depending on the reserve type.
Finally, \eqref{eqEquilibreZoneSud} represents the power balance in the south zone where the reserve requirement is modeled as an artificial load that needs to be supplied: The variable $\varPPMRsT{\ixRs}{\ixT}$ allows to detect if the generation is insufficient to deliver the reserve and forecasted load at the same time.

\scriptsize
\begin{align}
  & \begin{aligned}\label{eqEquilibreZoneSud}
    \paramPZT{\ixZ=\mathrm{south}}{\ixT} =
    &  \sum_{\ixChp \in \sChpZ{\ixZ}} \varPChpRsT{\ixChp}{\ixRs}{\ixT}
      + \sum_{\ixGp \in \sGpZ{\ixZ}} \varPGpRsT{\ixGp}{\ixRs}{\ixT}
      + \sum_{\ixIc \in \sIcZ{\ixZ}} \varPIcRsT{\ixIc}{\ixRs}{\ixT} \\
    & + \sum_{\ixIr \in \sIrZ{\ixZ}} \varPIrRsT{\ixIr}{\ixRs}{\ixT}
      + \sum_{\ixLi \in \sInLiZ{\ixZ}} \varPOutLiRsT{\ixLi}{\ixRs}{\ixT}
      - \sum_{\ixLi \in \sOutLiZ{\ixZ}} \varPInLiRsT{\ixLi}{\ixRs}{\ixT} \\
    & + \sum_{\ixRorhp \in \sRorhpZ{\ixZ}} \varPRorhpRsT{\ixRorhp}{\ixRs}{\ixT}
      + \varPOutMtdcZRsT{\ixZ}{\ixRs}{\ixT}
      - \varPInMtdcZRsT{\ixZ}{\ixRs}{\ixT} \\
    & + \varPResRT{\ixRs}{\ixT}
      + \varPPMRsT{\ixRs}{\ixT}
  \quad \forall \ixRs\;\forall \ixT
  \end{aligned}
\end{align}
\normalsize

\subsubsection{Objective function and detecting adequate level of reserve}

The power adequacy problem objective function is given by Equation \eqref{eqAlfredObj}, for each reserve $\ixRs$ and time step $\ixT$

\footnotesize
\begin{equation}
  \label{eqAlfredObj}
  \begin{aligned}
    \max \varPPMRsT{\ixRs}{\ixT}
  \end{aligned}
\end{equation}
\normalsize
The idea behind the objective function is the following:
If $\varPPMRsT{\ixRs}{\ixT}$ is greater than 0, then there is enough generation to deliver power to the load while respecting the reserve requirements;
otherwise, there is not enough generation to deliver power while respecting the reserve requirements, so that a reserve restoration algorithm needs to be run.

\subsection{Reserve restoration problem}

The restoration problem is fairly simple.
It relies on the same model as the power adequacy problem, except that it becomes a minimization problem.
For each problematic time step $\ixT$, an optimization problem that includes all reserves is built by adding a new constraint ensuring that $\varPPMRsT{\ixRs}{\ixT}$ is positive:

\footnotesize
\begin{equation}
  \label{eqAlfredPPM}
  \begin{aligned}
    \varPPMRsT{\ixRs}{\ixT} \geq 0, \ \forall \ixRs \forall \ixT
  \end{aligned}
\end{equation}
\normalsize

For each time step $\ixT$, the objective function becomes:

\footnotesize
\begin{equation}
  \label{eqAlfredCorrObj}
  \begin{aligned}
    \min \sum_{\ixRas \in \sRas} \paramRAS{\ixRas}{\ixT} \varbRAS{\ixRas}{\ixT}
  \end{aligned}
\end{equation}
\normalsize

The objective function minimize the remedial schemes needed by using the ones with a lower priority coefficient.
Due to space constraints, the implementation of each remedial action is not detailed, but it is worth mentioning that each remedial action results in modifying some variables or constraints of the model.
For instance, they can improve the stability limit's maximum value and allow more generation to reach the south zone, they can reduce the total load of the system or, for specific situations, abandon certain reliability constraints.
A non-exhaustive list of remedial actions includes (1) demand response, (2) tie line curtailment, (3) voltage reduction, (4) tap changer blocking, (5) use of emergency limits, (6) drop of reserve criteria and (7) load shedding. 

While using demand response programs or load shedding is common in power systems, HQ developed innovative remedial action by leveraging the specificity of its grid.
For example, since the majority of winter peak load is resistive heating, the voltage level of the transmission grid can be adjusted, reducing the load at the same time.
Similarly, during winter peak, the south zone is constrained by voltage collapse stability limits caused by automatic voltage control, done by the tap changer of the transformer.
HQ can remotely block tap changer regulation and improve the stability limits, allowing for more power to flow to the south zone.
The restoration problem returns a list of remedial action schemes.
Depending on whether the simulation is in real-time or in the operation planning horizon, the remedial action is either directly deployed or programmed into the system during the planning phase. 

\subsection{Implementation status}

\begin{table}
  \centering
  \caption{Some statistics about ALFRED in operation}\label{tab:stats-alfred}
  \begin{tabular}{llllll}
    \toprule                 
    Horizons     & TSD & ANTSH & RR  & TSACT   & ANTSE   \\
    \midrule                 
    Real-time    & 0s  & 1     & 30s & 15--20s & 120/h   \\
    0--4h        & 10m & 21    & 1m  & 25--30s & 1290/h  \\
    4--16h       & 30m & 24    & 2m  & 25--30s & 720/h   \\
    16--49h      & 60m & 33    & 5m  & 25--30s & 396/h   \\
    2d--10d      & 60m & 179   & 10m & 25--30s & 1077/h  \\
    10d--30d     & 60m & 480   & 20m & 25--30s & 1440/h  \\
    Current year & 24h & 168   & 24h & 25--30s & 168/24h \\
    Next year    & 24h & 365   & 24h & 25--30s & 365/24h \\
    \bottomrule
  \end{tabular}
  \vskip\baselineskip
  \begin{center}
    TSD: time step duration, ANTSH: average number of time steps per horizon, \\
    RR: refresh rate, TSACT: time step average computation time, \\
    ANTSE: average number of time step executions, \\
    s: seconds, m: minutes, h: hours
  \end{center}
\end{table}

ALFRED is already used by the operation teams at HQ.
It monitors reserve levels in real-time to a year from now.
\newtext{It replaced an old tool called GÉODE that did reserve monitoring.
GÉODE was using numerical values for stability limits that were extracted from the lookup tables described in Section~\ref{ss:simulations-required}.
It had no knowledge about what influenced the obtained values, so that after calculating the reserve levels, it needed to rescan the lookup table to obtain limits that were matching the simulation.
If the transits were exceeding these new limits, another iteration was required.
For performance reasons, the process could only be repeated a few times, leading in some cases to non stabilized, incomplete solutions.
If the result was negative, the reserve restoration problem needed even more iterations.
ALFRED, by explicitly integrating the stability constraints in the model, solves both the performance and the result quality issues.}

\newtext{
Table~\ref{tab:stats-alfred} provides some statistics about the ALFRED tool in operation, for different horizons.
The reserve monitoring calculations are refreshed more frequently and on smaller time steps as the time horizon moves closer to real-time.
For example, a 10-minute time step is used when monitoring the next four hours and is refreshed every 1 minute, while a daily time step is used for reserve level in a year, while being refreshed once a day.
On average, the computation time required is between 15 and 20 seconds for the real-time horizon, while it moves up to between 25 and 30 seconds in forecast horizons, whatever their duration.
}
It is also possible to import data into a study mode and analyze the specific consequences of a potential outage.


\section{RALPH: Solving the TSCUC}\label{sec:ralph}

This section shifts the focus to the tool used to solve the TSCUC problem at Hydro-Quebec, called RALPH\@.
In Subsection~\ref{ss:hydraulic}, additional concepts are introduced to facilitate the understanding of the TSUC model in a hydraulic context.
Subsection~\ref{ss:configs} describes a strategy to compute the set of hydro plant configurations that, in practice, significantly reduce the size of the problem.
Next, in Subsection \ref{ss:formulation} the TSCUC problem is detailed as a mixed integer linear program (MILP).
Finally, Subsection \ref{ss:report} reports preliminary results when solving the problem on real datasets.
The same notation convention as in Section~\ref{sec:alfred} is used for the indices, sets and relations (Table~\ref{tab:ralph-sets}), for the parameters and functions (Table~\ref{tab:ralph-parameters}), and for the decision variables and expressions (Table~\ref{tab:ralph-variables}).

\begin{table}[t]
  \caption{Indices, sets and relations}\label{tab:ralph-sets}
  \scriptsize
  \centering
  \begin{tabular}{>{\centering\arraybackslash}r@{\hspace{1mm}}c@{\hspace{1mm}}lm{6.0cm}}
    \toprule
    Index & & Set & Description \\
    \midrule
    $\ixCfg$ & $\in$ & $\sCfg$
      & Set of all generator configurations \\
    $\ixDh$ & $\in$ & $\sDh$
      & Set of all drop heights ($\dhLow$, $\dhHi$) \\
    $\ixG$ & $\in$ & $\sG$
      & Set of all hydro generators \\
    $\ixG$ & $\in$ & $\sGNed$
      & Set of all hydro generators not participating to the economic dispatch \\
    $\ixHp$ & $\in$ & $\sHp$
      & Set of all (controllable) hydro plants \\
    $\ixRe$ & $\in$ & $\sRe$
      & Set of all reservoirs \\
    $\ixRi$ & $\in$ & $\sRi$
      & Set of all rivers \\
    $\ixSp$ & $\in$ & $\sSp$
      & Set of all spillways \\
    $\ixSz$ & $\in$ & $\sSz$
      & Set of all stability zones \\
    $\ixY$ & $\in$ & $\sY$
      & Set of all normal generator yields ($\yMin$, $\yOpt$, $\yMax$) \\
    $\ixY$ & $\in$ & $\sYStab$
      & Set of all generator yields ($\yMin$, $\yOpt$, $\yMax$, $\yStab$) \\
    $\ixZ$ & $\in$ & $\sZ$
      & Set of all zones \\
    \bottomrule
  \end{tabular}
  \vskip\baselineskip
  \begin{tabular}{m{0.5cm}m{7cm}}
    \toprule
    Set & Description \\
    \midrule
    $\mTFcpl{\ixFcpl}$
      & Time step at which FCPL $\ixFcpl$ is active \\
    $\sCfgG{\ixG}$
      & Set of all configurations involving generator $\ixG$ \\
    $\sCfgHp{\ixHp}$
      & Set of all configurations of hydro plant $\ixHp$ \\
    $\sCfgSg{\ixSg}$
      & Set of all configurations involving super-generator $\ixSg$ \\
    $\sHpRe{\ixRe}$
      & Set of hydro plants supplied by reservoir $\ixRe$ \\
    $\sHpRi{\ixRi}$
      & Set of hydro plants discharging into river $\ixRi$ \\
    $\sHpSz{\ixSz}$
      & Set of hydro plants in stability zone $\ixSz$ \\
    $\sRiRe{\ixRe}$
      & Set of rivers ending in reservoir $\ixRe$ \\
    $\mSgGT{\ixG}{\ixT}$
      & Super-generator containing generator $\ixG$ during time step $\ixT$ \\
    $\mTSg{\ixSg}$
      & Time step of super-generator $\ixSg$ \\
    $\mReSg{\ixSg}$
      & Reservoir associated with super-generator $\ixSg$ \\
    $\sSgT{\ixT}$
      & Set of all super-generators active at time step $\ixT$ \\
    $\sSpRe{\ixRe}$
      & Set of spillways available to reservoir $\ixRe$ \\
    $\sSpRi{\ixRi}$
      & Set of spillways discharging into river $\ixRi$ \\
    \bottomrule
  \end{tabular}
\end{table}

\begin{table}[t]
  \caption{Parameters and functions}\label{tab:ralph-parameters}
  \scriptsize
  \centering
  \begin{tabular}{m{1.8cm}m{6.0cm}}
    \toprule
    Parameter & Description \\
    \midrule
    $\paramDurationT{\ixT}$
      & Duration of time step $\ixT$ \seco{} \\
    \midrule
    $\paramLamRiTTp{\ixRi}{\ixT}{\ixTp}$
      & Proportion of volume whose transit between
        time steps $\ixT$ and $\ixTp$ in river $\ixRi$ 
        is completed \\
    \midrule
    $\paramPGCfgYDh{\ixG}{\ixCfg}{\ixY}{\ixDh}$
      & Power of generator $\ixG$ at yield $\ixY$ and
        drop height $\ixDh$ when configuration $\ixCfg$
        is active \MW{} \\
    $\paramPSgCfgYDh{\ixSg}{\ixCfg}{\ixY}{\ixDh}$
      & Power of super-generator $\ixSg$ at yield $\ixY$
        and drop height $\ixDh$ when configuration $\ixCfg$ is active \MW{} \\
    $\paramSfcDownSgCfgYDh{\ixSg}{\ixCfg}{\ixY}{\ixDh}$/
    $\paramSfcUpSgCfgYDh{\ixSg}{\ixCfg}{\ixY}{\ixDh}$
      & Available power decrease/increase of super-generator $\ixSg$ at yield
        $\ixY$ and drop height $\ixDh$ when configuration $\ixCfg$ is active
        ($0$ if $\ixSg$ does not contribute to SFC) \MW{} \\
    $\paramUpperStabSgCfgYDh{\ixSg}{\ixCfg}{\ixY}{\ixDh}$
      & Maximum power output margin of super-generator $\ixG$ at yield $\ixY$
        and drop height $\ixDh$ when configuration $\ixCfg$ is active during
        transient events \MW{} \\
    \midrule
    $\paramFGCfgYDh{\ixG}{\ixCfg}{\ixY}{\ixDh}$
      & Flow turbined by generator $\ixG$ at yield $\ixY$ and
        drop height $\ixDh$ when configuration $\ixCfg$ is active \mcps{} \\
    \midrule
    $\paramVInitRe{\ixRe}$
      & Initial volume of reservoir $\ixRe$ \mc{} \\
    \midrule
    $\paramLReDh{\ixRe}{\ixDh}$
      & Level of reservoir $\ixRe$ at drop height $\ixDh$ \meter{} \\
    \midrule
    $\paramSfcDown$/
    $\paramSfcUp$/
    $\paramSfcTot$
      & Low/high/total range threshold for secondary frequency control \MW{} \\
    $\paramStabAbsSz{\ixSz}$
      & Minimum stability reserve threshold to maintain in stability zone
        $\ixSz$ \MW{} \\
    $\paramStabRateSz{\ixSz}$
      & Rate of stability power to maintain in stability zone $\ixSz$ \MW{} \\
    \midrule
    $\costManCfgT{\ixCfg}{\ixT}$
      & Cost of changing to/from configuration $\ixCfg$ at time step $\ixT$
        \doll{} \\
    $\costSetGT{\ixG}{\ixT}$
      & Cost of changing the setpoint of generator $\ixG$ at time
        step $\ixT$ \dpmw{} \\
    \midrule
    $\fctVLRe{\ixRe}{\cdot}$
      & Linear function approximating the level of reservoir $\ixRe$ from its
        volume \mcpstm{} \\
    \midrule
    $\fctPfcLimitT{\ixT}{\cdot}$
      & Piecewise linear function computing a stability threshold at time step
        $\ixT$ from the worst north and south FCPL \MWMWtMW{} \\
    \midrule
    $\fctUpperNorthFcpl{\cdot}$
      & Piecewise linear function expressing the north FCPL limit with respect
        to the total maximum output power at time step $\ixT$ \MWMWtMW{} \\
    \bottomrule
  \end{tabular}
\end{table}

\begin{table}[t]
  \caption{Decision variables and expressions}\label{tab:ralph-variables}
  \scriptsize
  \centering
  \begin{tabular}{m{1.2cm}m{6.4cm}}
    \bottomrule
    Variable & Description \\
    \midrule
    $\varACfgT{\ixCfg}{\ixT}$
      & Indicator if configuration $\ixCfg$ is active during time step $\ixT$ \\
    $\varAGT{\ixG}{\ixT}$
      & Indicator if generator $\ixG$ is active during time step $\ixT$ \\
    $\varHSg{\ixSg}$
      & Indicator if super-generator $\ixSg$ produces with high yield,
        \textit{i.e.} yield between $\yOpt$ and $\yMax$ \\
    \midrule
    $\varWSgCfgYDh{\ixSg}{\ixCfg}{\ixY}{\ixDh}$
      & Production weight of super-generator $\ixSg$ at yield $\ixY$ and
        drop height $\ixDh$ when configuration $\ixCfg$ is active \\
    \midrule
    $\varPGT{\ixG}{\ixT}$/
    $\varPHpT{\ixHp}{\ixT}$
      & Power produced by generator $\ixG$/hydro plant $\ixHp$ during time step
        $\ixT$ \MW{} \\
    $\varSfcDownHpT{\ixHp}{\ixT}$/
    $\varSfcUpHpT{\ixHp}{\ixT}$
      & Available power decrease/increase of hydro plant $\ixHp$ during time step
        $\ixT$ \MW{} \\
    $\varPSgY{\ixSg}{\ixY}$
      & Power produced by super-generator $\ixSg$ at yield $\ixY$ \MW{} \\
    $\varPTransSg{\ixSg}$
      & Maximum output power of super-generator $\ixSg$ during transient events
        \MW{} \\
    $\varPTransT{\ixT}$
      & Maximum total output power during transient events at time step $\ixT$
        \MW{} \\
    $\varPWorstT{\ixT}$
      & Production of the worst FCPL during time step $\ixT$ \\
    $\varPSg{\ixSg}$
      & Power produced by super-generator $\ixSg$ \MW{} \\
    $\varSfcDownSg{\ixSg}$/
    $\varSfcUpSg{\ixSg}$
      & Available power decrease/increase of super-generator $\ixSg$ \MW{} \\
    $\varPPfcSg{\ixSg}$
      & Available PFC control margin of super-generator $\ixSg$ \MW{} \\
    \midrule
    $\varFHpT{\ixHp}{\ixT}$
      & Flow passing through hydro plant $\ixHp$ during time step $\ixT$ \mcps{}
        \\
    \midrule
    $\varLReT{\ixRe}{\ixT}$
      & Average level of reservoir $\ixRe$ during time step $\ixT$ \meter{} \\
    \midrule
    $\varVReT{\ixRe}{\ixT}$
      & Volume in reservoir $\ixRe$ at the end of time step $\ixT$ \mc{} \\
    $\varVInRiT{\ixRi}{\ixT}$/
    $\varVOutRiT{\ixRi}{\ixT}$
      & Volume entering/leaving river $\ixRi$ during time step $\ixT$ \mc{} \\
    $\varVSpT{\ixSp}{\ixT}$
      & Volume passing through spillway $\ixSp$ during time step $\ixT$ \mc{} \\
    \bottomrule
  \end{tabular}
  \vskip\baselineskip
  \begin{tabular}{m{1.4cm}m{6.2cm}}
    \toprule
    Expression & Description \\
    \midrule
    $\exprLowerLimitMtdcT{\ixT}$
      & Piecewise linear expression representing the south
        FCPL to cover at time step $\ixT$ \MW{} \\
    \midrule
    $\exprUpperLimitSc{\ixSc}$
      & Piecewise linear expression representing the upper
        limit of topology constraint $\ixTc$ \MW{} \\
    \bottomrule
  \end{tabular}
\end{table}

\subsection{Hydraulicity}\label{ss:hydraulic}


\newtext{
In order to manage its hydraulicity in a consistent manner, HQ divides its hydro plants in two categories: (1) non-controllable hydro plants, which are run-of-river hydro plants or hydro plants with small reservoirs, and (2) controllable hydro plants, which are hydro plants with large reservoirs that must be managed with a long-term view.
In particular, the power produced by non-controllable hydro plants is considered predetermined for the RALPH tool, so that only the decisions about controllable hydro plants are optimized.
}

The water transiting along a river does not all travel at the same speed.
To take into account this reality, a \emph{\newtext{unit hydrograph}} $\Phi_{\ixRi}: \R^+ \rightarrow [0,1]$ is associated with each river $\ixRi$, mapping an elapsed time $t$ onto the proportion $\Phi_{\ixRi}(t)$ of water that has completed its transit along river $\ixRi$.
To ensure water volume conservation, the function satisfies the identity
$\int_{t \in \R^+} \Phi_{\ixRi}(t)dt = 1$.
In practice, the exact shape of $\Phi_{\ixRi}$ is unknown, but can be approximated by computing coefficients $\paramLamRiTTp{\ixRi}{\ixT}{\ixTp}$ estimating the proportion of water whose transit between time steps $\ixT$ and $\ixTp$ in river $\ixRi$ is completed.
\newtext{Unit hydrograph calculations used by HQ are detailed in \cite{bisson1983}. They have been used over the years in multiple applications like climate change impact assessment~\cite{minville2010} and long-term water management strategies using stochastic dynamic programming~\cite{cote2011}. More recently, new methods to compute unit hydrographs have been proposed using linear programming and convolutional neural network. \cite{Feinstein2023}.}

\begin{figure}[t]
  \centering
  \includegraphics[width=.98\linewidth]{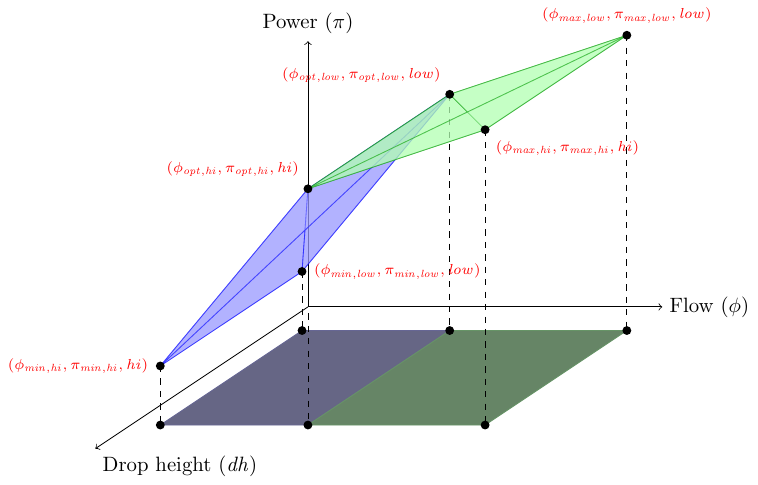}
  \caption{The two tetrahedra in which the production weights must be chosen}\label{fig:tetrahedra}
\end{figure}

The power produced by a hydro generator is given by
\begin{equation}\label{eq:phq}
  P(H, Q) = \rho \cdot g \cdot Q \cdot H \cdot \eta(Q)
\end{equation}
where $H$ is the drop height (in meter), $Q$ is the turbined flow (in m$^3$/s), $\rho \approx 997$ is the volumic mass of water (in kg/m$^3$), $g \approx 9.81$ is the gravity acceleration (in m/s$^2$) and $\eta(Q) < 1$ is the generator yield when the flow is $Q$ \cite{ginocchio2012energie}.
For a given generator, the function $\eta$ can be complex.
Moreover, Equation \eqref{eq:phq} is nonlinear.
Therefore, to represent the problem in the MILP context, the realizable surface is approximated by a piecewise linear function, whose geometric representation consists of two tetrahedra (see Figure \ref{fig:tetrahedra}).
The vertices of those tetrahedra are computed with the help of an internal tool called SChOp, whose main function is to solve the economic dispatch problem by retrieving the real-time hydraulic characteristics of hydro plant generators.
The production linearization is articulated around three normal \emph{yields}, according to if the generator produces at its minimum  ($\yMin$), its optimal ($\yOpt$) or its maximum ($\yMax$) yield.
There is also a special yield, labeled $\yStab$ at which generators can produce above the maximum yield $\yMax$ for a short period of time in case of a stability event.
\newtext{
This linear approximation is considered acceptable for operational purposes, since
(1) the tetrahedra of Figure~\ref{fig:tetrahedra} are in practice almost planar, so that the error resulting of not being perfectly on the curve plan in the middle of it is very small,
(2) the time horizon is short with respect to the large reservoir capacity of controllable hydro plants and
(3) the refresh rate of regenerating a production plan is high enough to benefit from frequently updated SCADA to ensure that the errors are not cumulative.
}

Furthermore, hydro generators that present similar characteristics and operational constraints should produce energy at the same yield.
To avoid introducing additional complexity to the mathematical model while cleanly taking into account this reality, it is convenient to introduce the notion of \emph{super-generator}.
Roughly speaking, a super-generator is a set of generators whose production at a given time step should be processed in a similar manner.
The set of super-generators at a given time step can be computed by using an \emph{apartness relation} $\apart$ defined as follows.
Let $\ixG$ and $\ixGd$ be two generators.
Then $\ixG$ and $\ixGd$ are distinguished (or apart), written $\ixG \apart \ixGd$, if at least one of the following conditions is verified:
(1) $\ixG$ and $\ixGd$ belong to distinct hydro plants, (2) either $\ixG$ or $\ixGd$ is subject to a specific operational restrictions, (3) $\ixG$ and $\ixGd$ are separated by at least one FCPL set or (4) $\ixG$ and $\ixGd$ are separated by at least one topology constraint.
Otherwise, $\ixG$ and $\ixGd$ are said to be indistinguishable, or equivalent.
A \emph{super-generator} is then simply a set of equivalent generators at a given time step.

\subsection{Generation of hydro plant configurations}\label{ss:configs}

The production of a hydro generator depends on the state of the upstream and downstream levels.
Also, it might depend on the state of the other generators in the same hydro plant, since some generators share upstream water pipes, or convey water to the same downstream location.
In order to faithfully represent this observation, the concept of \emph{hydro plant configurations}, i.e. subsets of generators belonging to some given hydro plant $\ixHp$, is introduced.
In theory, there can be up to $2^{|\sGHp{\ixHp}|}$ possible configurations, but in practice, many of those configurations need not be considered if only the operational constraints are taken into account.
More precisely, each generator can be assigned a \emph{commitment order} denoted by $\paramcommitOrderg{\ixG}$ prioritizing it according to (1) its forced synchronization, (2) its availability, (3) its power restrictions, as well as (4) the FCPL constraints, (5) the topology constraints it is subjected to and (6) other mechanical differences between generators that can impact their yield.
This commitment order can then be used to generate a minimal set of hydro plant configurations that needs to be considered in the MILP formulation.

\begin{algorithm}[t]
\small
  \caption{Generation of hydro plant configurations}
  \label{algorithm-configs-generation}
  \begin{algorithmic}[1]
    \Function{Configs}{$\ixHp$ : hydro plant}
      \State $\sCfgHp{\ixHp} \gets \emptyset$
      \State $\algforcedgroups \gets \{\ixG \in \sGHp{\ixHp} \mid \mbox{$\ixG$ is always forced online}\}$ \label{config-alg:init-forced}
      \State $\algunavailgroups \gets \{\ixG \in \sGHp{\ixHp} \mid \mbox{$\ixG$ is unavailable on $\sT$}\}$ \label{config-alg:init-unavail}
      \State $\Call{GenerateConfigs}{\sCfgHp{\ixHp}, \algforcedgroups, \algunavailgroups \cup \algforcedgroups}$ \label{config-alg:init-call}
      \State \Return $\sCfgHp{\ixHp}$
    \EndFunction
      
	\Procedure{GenerateConfigs}{$\sCfgHp{\ixHp}, \algconfiggroups, \algunavailgroups$}
		  \State $\algtoaddgroups \gets \emptyset$
		  \State $co_{\mathrm{max}} \gets \max\{\paramcommitOrderg{\ixG} \mid \ixG \in \algconfiggroups \setminus \algunavailgroups\}$
      \State $\algpotentialgroups \gets \{\ixG \in \sGHp{\ixHp} \setminus \algunavailgroups \mid \paramcommitOrderg{\ixG} > co_{max}\}$\label{config-alg:candidates}
      \State $\algtoaddgroups \gets \algtoaddgroups \cup \{\ixG \in \algpotentialgroups \mid \mbox{$\ixG$ is forced online at least once}\}$ \label{config-alg:forced-once}
        \State $\algtoaddgroups \gets \algtoaddgroups \cup \Call{AvailabilitiesCover}{\algpotentialgroups}$ \label{config-alg:step10}
      \State $\algtoaddgroups \gets \algtoaddgroups \cup \Call{RestrictionsCover}{\algpotentialgroups}$ \label{config-alg:step11}
      \State $\algtoaddgroups \gets \algtoaddgroups \cup \Call{ConstraintsCover}{\algpotentialgroups}$ \label{config-alg:step12}
		  \For{$g \in \algtoaddgroups$} \label{config-alg:step13}
        \State \newtext{$\algconfiggroups' \gets \algconfiggroups \cup \{\ixG\}$}
        \State \newtext{$\sCfgHp{\ixHp} \gets \sCfgHp{\ixHp} \cup \{\algconfiggroups'\}$ \label{config-alg:step14}}
        \State \newtext{\Call{GenerateConfigs}{$\sCfgHp{\ixHp}, \algconfiggroups', \algunavailgroups$} \label{config-alg:step15}}
		  \EndFor
	\EndProcedure
\end{algorithmic}
\end{algorithm}

Algorithm~\ref{algorithm-configs-generation} describes in more detail the generation of the configurations set $\sCfgHp{\ixHp}$ for a given controllable hydro plant $\ixHp \in \sHp$.
The algorithm starts by identifying the generators that are forced synchronized during the entire time horizon (Line~\ref{config-alg:init-forced}) as they must be part of every configuration and thus constitute the initial one.
On the opposite side, the unavailable generators can be ignored (Line~\ref{config-alg:init-unavail}).
The algorithm then proceeds by calling the recursive procedure $\Call{GenerateConfigs}{}$ (Line~\ref{config-alg:init-call}).

The procedure $\Call{GenerateConfigs}{}$ aims to generate configurations by extending the current configuration with generators that cover missing availabilities, restrictions, FCPL, or topology constraints.
As a first step, a set of candidate generators is computed (Line \ref{config-alg:candidates}).
Then the candidate generators that are forced online at least once during the time horizon are added, since they must be part of all configuration extensions (Line~\ref{config-alg:forced-once}).
The next three steps (Lines~\ref{config-alg:step10}-\ref{config-alg:step12}) aim to add as few as possible generators in a way such that they cover as much as possible availability, unrestricted production and unconstrained FCPL/topological sets of generators.
Due to space restriction, the $\Call{AvailabilitiesCover}{}$, $\Call{RestrictionsCover}{}$ and $\Call{ConstraintsCover}{}$ functions are not detailed here.
Also, it is worth mentioning that those three steps are independent and can be performed in any order.
To process the availabilities cover, the idea consists of ensuring that there are enough generators that are available at each time step.
Hence, if the first generator is available at all time steps, a single generator is added, otherwise, a subset of the candidate generators is selected such that a maximum number of time steps is covered.
Restrictions, FCPL, and topology constraints are handled similarly by favoring a minimum number of active restrictions/constraints over the time horizon.
Since candidate generators are sorted by their commitment order, which prioritizes the available generators with no restrictions/constraints above all others, the number of generators to add is minimized.
This is why the inclusion of a new generator is done only at the end of the procedure (Lines~\ref{config-alg:step13}-\ref{config-alg:step15}).
In the ideal case where all the generators in the hydro plant are available, with no active constraints or restrictions, the algorithm generates one configuration per number of generators.
On real-case scenarios, most of the time, the ideal case is the norm.
Hence, in practice, Algorithm~\ref{algorithm-configs-generation} provides a robust solution for controlling the theoretical combinatorial explosion of the hydro plant configuration generation, especially for the hydro plants that contain many generators.

\subsection{Mathematical formulation}\label{ss:formulation}

The next section is devoted to the detailed description of the underlying mathematical model.

\subsubsection{Production}\label{sss:production}

Water being the main source of energy, constraints must be defined to ensure that it travels through the various components (hydro plants, reservoirs, rivers, and spillways) consistently:

\footnotesize
\begin{align}
  & \begin{aligned}
    \varVReT{\ixRe}{\ixTz} & =
      \paramVInitRe{\ixRe}
  & \forall \ixRe\ \\
  \end{aligned} \\
  & \begin{aligned}
    \varVReT{\ixRe}{\ixT} & =
      \varVReT{\ixRe}{\ixT-1}
      + \sum_{\ixRi \in \sRiRe{\ixRe}} \varVOutRiT{\ixRi}{\ixT} \\
      & - \sum_{\ixHp \in \sHpRe{\ixRe}} \paramDurationT{\ixT} \varFHpT{\ixHp}{\ixT}
      - \sum_{\ixSp \in \sSpRe{\ixRe}} \varVSpT{\ixSp}{\ixT}
  & \forall \ixRe\;\forall \ixT \in \sTf \\
  \end{aligned} \\
  & \begin{aligned}
    \varLReT{\ixRe}{\ixT} & =
      \fctVLRe{\ixRe}{(\varVReT{\ixRe}{\ixT-1} + \varVReT{\ixRe}{\ixT})/2}
  & \forall \ixRe\;\forall \ixT \in \sTf \\
  \end{aligned} \\
  & \begin{aligned}
    \varVInRiT{\ixRi}{\ixT} & =
      \sum_{\ixHp \in \sHpRi{\ixRi}} \paramDurationT{\ixT} \varFHpT{\ixHp}{\ixT}
      + \sum_{\ixSp \in \sSpRi{\ixRi}} \varVSpT{\ixSp}{\ixT}
  & \forall \ixRi\;\forall \ixT \\
  \end{aligned} \\
  & \begin{aligned}
    \varVOutRiT{\ixRi}{\ixT} & =
      \sum_{\ixTp \leq \ixT} \paramLamRiTTp{\ixRi}{\ixTp}{\ixT} \varVInRiT{\ixRi}{\ixTp}
  & \forall \ixRi\;\forall \ixT \\
  \end{aligned}
\end{align}
\normalsize

The core of the unit commitment problem is expressed through the following constraints, where the production weights $\varWSgCfgYDh{\ixSg}{\ixCfg}{\ixY}{\ixDh}$ must be selected:

\footnotesize
\begin{align}
  & \begin{aligned}
    \varHSg{\ixSg} & \geq
      \sum_{\ixCfg \in \sCfgSg{\ixSg}}
        \sum_{\ixDh \in \sDh}
          \varWSgCfgYDh{\ixSg}{\ixCfg}{\yMax}{\ixDh}
  & \forall \ixSg\ \\
  \end{aligned} \\
  & \begin{aligned}
    1 - \varHSg{\ixSg} & \geq
      \sum_{\ixCfg \in \sCfgSg{\ixSg}}
        \sum_{\ixDh \in \sDh}
          \varWSgCfgYDh{\ixSg}{\ixCfg}{\yMin}{\ixDh}
  & \forall \ixSg\ \\
  \end{aligned} \\
  & \begin{aligned}
    1 & =
      \sum_{\ixCfg \in \sCfgHp{\ixHp}} \varACfgT{\ixCfg}{\ixT}
  \end{aligned} \\
  & \begin{aligned}
    \varACfgT{\ixCfg}{\mTSg{\ixSg}} & =
      \sum_{\ixY \in \sY}
        \sum_{\ixDh \in \sDh}
          \varWSgCfgYDh{\ixSg}{\ixCfg}{\ixY}{\ixDh}
  & \forall \ixCfg\;\forall \ixSg \\
  \end{aligned} \\
  & \begin{aligned}
    \varAGT{\ixG}{\ixT} & =
      \sum_{\ixCfg \in \sCfgG{\ixG}} \varACfgT{\ixCfg}{\ixT}
  & \forall \ixG\;\forall \ixT \\
  \end{aligned}
\end{align}
\normalsize

Once the units are committed, the powers of the hydro plant configurations, generators, hydro plants, and super-generators can be retrieved, as well as the flows turbinated by the hydro plants and the level variations of the reservoirs:

\footnotesize
\begin{align}
  & \begin{aligned}
    \varPHpT{\ixHp}{\ixT} & =
      \sum_{\ixG \in \sGHp{\ixHp}} \varPGT{\ixG}{\ixT}
  & \forall \ixHp\;\forall \ixT \\
  \end{aligned} \\
  & \begin{aligned}
    \varPGT{\ixG}{\ixT} & =
      \sum_{\ixCfg \in \sCfgG{\ixG}}
        \sum_{\ixY \in \sY}
          \sum_{\ixDh \in \sDh}
            \paramPGCfgYDh{\ixG}{\ixCfg}{\ixY}{\ixDh}
            \ \varWSgCfgYDh{\mSgGT{\ixG}{\ixT}}{\ixCfg}{\ixY}{\ixDh}
  & \forall \ixG\;\forall \ixT \\
  \end{aligned} \\
  & \begin{aligned}
    \varPSgY{\ixSg}{\ixY} & =
      \sum_{\ixCfg \in \sCfgSg{\ixSg}}
        \sum_{\ixDh \in \sDh}
          \paramPSgCfgYDh{\ixSg}{\ixCfg}{\ixY}{\ixDh}
          \ \varWSgCfgYDh{\ixSg}{\ixCfg}{\ixY}{\ixDh}
  & \forall \ixSg\;\forall \ixY \\
  \end{aligned} \\
  & \begin{aligned}
    \varPSg{\ixSg} & =
      \sum_{\ixY \in \sY}
        \varPSgY{\ixSg}{\ixY}
  & \forall \ixSg \\
  \end{aligned} \\
  & \begin{aligned}
    \varFHpT{\ixHp}{\ixT} & =
      \sum_{\ixG \in \sGHp{\ixHp}}
        \sum_{\ixCfg \in \sCfgSg{\ixSg}}
          \sum_{\ixY \in \sY}
            \sum_{\ixDh \in \sDh}
              \paramFGCfgYDh{\ixG}{\ixCfg}{\ixY}{\ixDh}
              \ \varWSgCfgYDh{\mSgGT{\ixG}{\ixT}}{\ixCfg}{\ixY}{\ixDh}
  & \forall \ixHp\;\forall \ixT \\
  \end{aligned} \\
  & \begin{aligned}
    \varLReT{\mReSg{\ixSg}}{\mTSg{\ixSg}} & =
      \sum_{\ixCfg \in \sCfgSg{\ixSg}}
        \sum_{\ixY \in \sY}
          \sum_{\ixDh \in \sDh}
            \paramLReDh{\mReSg{\ixSg}}{\ixDh}
            \ \varWSgCfgYDh{\ixSg}{\ixCfg}{\ixY}{\ixDh}
  & \forall \ixSg\ \\
  \end{aligned}
\end{align}
\normalsize

\subsubsection{Frequency containment constraints}\label{sss:stability-limits}

ALFRED's reserves calculations ensure that enough resources are available to generate a UC that satisfies minimal primary frequency control (PFC) margins and secondary frequency control (SFC) margins criteria.
However, the UC itself needs to modelize those constraints more precisely to provide a solution that really respects those minimal criteria:

\footnotesize
\begin{align}
  & \begin{aligned}
    \label{eqSFCUpSG}
    \varSfcUpSg{\ixSg} & =
      \sum_{\ixCfg \in \sCfgSg{\ixSg}}
        \sum_{\ixY \in \sY}
          \sum_{\ixDh \in \sDh}
            \paramSfcUpSgCfgYDh{\ixSg}{\ixCfg}{\ixY}{\ixDh}
            \ \varWSgCfgYDh{\ixSg}{\ixCfg}{\ixY}{\ixDh}
  & \forall \ixSg\ \\
  \end{aligned} \\
  & \begin{aligned}
    \label{eqSFCDnSG}
    \varSfcDownSg{\ixSg} & =
      \sum_{\ixCfg \in \sCfgSg{\ixSg}}
        \sum_{\ixY \in \sY}
          \sum_{\ixDh \in \sDh}
            \paramSfcDownSgCfgYDh{\ixSg}{\ixCfg}{\ixY}{\ixDh}
            \ \varWSgCfgYDh{\ixSg}{\ixCfg}{\ixY}{\ixDh}
  & \forall \ixSg\ \\
  \end{aligned} \\
  & \begin{aligned}
    \label{eqSFCUpHP}
    \varSfcUpHpT{\ixHp}{\ixT} & =
      \sum_{\ixSg \in \sSgHpT{\ixHp}{\ixT}}
        \varSfcUpSg{\ixSg}
  & \forall \ixHp\;\forall \ixT\ \\
  \end{aligned} \\
  & \begin{aligned}
    \label{eqSFCDnHP}
    \varSfcDownHpT{\ixHp}{\ixT} & =
      \sum_{\ixSg \in \sSgHpT{\ixHp}{\ixT}}
        \varSfcDownSg{\ixSg}
  & \forall \ixHp\;\forall \ixT\ \\
  \end{aligned} \\
  & \begin{aligned}
    \label{eqSFCUp}
    \sum_{\ixHp \in \sHp} \varSfcUpHpT{\ixHp}{\ixT} & \geq
      \paramSfcUp
  & \forall \ixT\ \\
  \end{aligned} \\
  & \begin{aligned}
    \label{eqSFCDn}
    \sum_{\ixHp \in \sHp} \varSfcDownHpT{\ixHp}{\ixT} & \geq
      \paramSfcDown
  & \forall \ixT\ \\
  \end{aligned} \\
  & \begin{aligned}
    \label{eqSFCTotal}
    \sum_{\ixHp \in \sHp}
      \left(\varSfcDownHpT{\ixHp}{\ixT} + \varSfcUpHpT{\ixHp}{\ixT}\right) & \geq
        \paramSfcTot
  & \forall \ixT\ \\
  \end{aligned}
\end{align}
\normalsize

Equations~\eqref{eqSFCUpSG}--\eqref{eqSFCDnHP} retrieve the SFC up and down regulation margins for each super-generator and hydro plant, while Equations~\eqref{eqSFCUp}--\eqref{eqSFCTotal} ensure that there is enough SFC up, down and total regulation margins for the whole grid.
Equations~\eqref{eqPEffStabSG}--\eqref{eqPEffStabTotal} compute the maximum output of all generators during a frequency event, which is then used as an approximation of the system strength and inertia.
This can be used to determine the maximum value of worst FCPL and the required PFC margins:

\footnotesize
\begin{align}
  & \begin{aligned}
    \label{eqPEffStabSG}
    \varPTransSg{\ixSg}
    & = \sum_{\ixCfg \in \sCfgSg{\ixSg}}
          \sum_{\ixDh \in \sDh}
            \sum_{\ixY \in \sY}
              \paramPSgCfgYDh{\ixSg}{\ixCfg}{\yStab}{\ixDh}
              \ \varWSgCfgYDh{\ixSg}{\ixCfg}{\ixY}{\ixDh}
  & \forall \ixSg \\
  \end{aligned} \\
  & \begin{aligned}
    \label{eqPEffStabTotal}
    \varPTransT{\ixT}
    & = \sum_{\ixSg \in \sSgT{\ixT}} \varPTransSg{\ixSg}
  & \forall \ixT\ \\
  \end{aligned}
\end{align}
\normalsize

PFC margins are tied to the worst FCPL\@.
However, since identifying a second worse FCPL is not necessary, the model for FCPL identification is simpler.
Equations~\eqref{eqLimiteUBFCPLRalph}--\eqref{eqLimiteLBFCPLNorth} bound the worst FCPL\@:

\footnotesize
\begin{align}
  & \begin{aligned}
     \label{eqLimiteUBFCPLRalph}
     \varPWorstT{\ixT}
     & \leq \fctUpperNorthFcpl{\varPTransT{\ixT}}
   & \forall \ixT\ \\
   \end{aligned} \\
  & \begin{aligned}
     \label{eqLimiteLBFCPLNorth}
     \varPWorstT{\ixT}
     & \geq \sum_{\ixSg \in \sFcplSg{\ixFcpl}}
       \left(\varPSg{\ixSg} + \varSfcUpSg{\ixSg}\right)
   & \forall \ixFcpl\ \\
   \end{aligned} \\
  & \begin{aligned}
    \label{eqPFCSG}
    \varPPfcSg{\ixSg}
    & = \sum_{\ixCfg \in \sCfgSg{\ixSg}}
      \sum_{\ixY \in \sY}
        \sum_{\ixDh \in \sDh}
          \paramUpperStabSgCfgYDh{\ixSg}{\ixCfg}{\ixY}{\ixDh}
          \ \varWSgCfgYDh{\ixSg}{\ixCfg}{\ixY}{\ixDh}
  & \forall \ixSg\ \\
  \end{aligned} \\
  & \begin{aligned}
    \label{eqPFCTotal}
    \sum_{\ixSg \in \sSgT{\ixT}}
      \varPPfcSg{\ixSg}
    & \geq 
            \fctPfcLimitT{\ixT}{\varPWorstT{\ixT}}
  & \forall \ixT\ \\
  \end{aligned} \\
  & \begin{aligned}
    \label{eqPFCZone}
    \sum_{\ixHp \in \sHpSz{\ixSz}}
      \sum_{\ixSg \in \sSgHpT{\ixHp}{\ixT}}
        \varPPfcSg{\ixSg}
    & \geq
      \min(\paramStabAbsSz{\ixSz},
           \paramStabRateSz{\ixSz}\varPTransT{\ixT})
  & \forall \ixSz\;\forall \ixT \\
  \end{aligned}
\end{align}
\normalsize

Equation~\eqref{eqPFCSG} returns the PFC regulation margin of each super generator.
Equation~\eqref{eqPFCTotal} ensures sufficient PFC regulation margins for the whole system.
Equation~\eqref{eqPFCZone} adds PFC constraints to each zone.

\subsubsection{Transmission}\label{sss:transmission}

The transmission model of RALPH is similar to the one of ALFRED\@.
Equations~\eqref{eqPertesLienAlfred}--\eqref{eqLimite2MtdcAlfred} are repeated to describe the power flow between zones, for each time step.
Equation~\eqref{eqEquilibreZone1} ensures that the power is balanced in every zone, including the south one, since the HTSCUC problem aims to balance generation and load instead of determining the maximum capacity of the grid.
Reserve constraints are omitted since ALFRED already ensures an adequate level.
Equation~\eqref{eqContrainteEnsembleRalph} has the same purpose as Equation~\eqref{eqContrainteEnsembleAlfred} but relies on the super-generator concept introduced in the HTSCUC formulation.
The SFC up margin is also taken into account to consider the worst-case scenario:

\footnotesize
\begin{align}
  \label{eqContrainteEnsembleRalph}
  & \begin{aligned}
      \sum_{\ixSg \in \sTcSg{\ixTc}}
        \left(\varPSg{\ixSg} + \varSfcUpSg{\ixSg}\right)
      & \leq \exprUpperLimitTc{\ixTc}
  & \forall \ixTc\ \\
  \end{aligned}
\end{align}
\normalsize

\subsubsection{Objective function}\label{sss:objective}

\newtext{
In contrast with the classical hydro unit commitment problem which aims to optimize the yield of hydro generators~\cite{Anjos2017}, HQ considers a significantly different objective function that aims to correctly manage its controllable hydro plants with a long-term view.
First, many hydraulic considerations are encoded in hard and soft constraints on reservoir levels that can be communicated by the operators through level bounds, taking into account the stochasticity of the natural water inflows brought to the reservoirs.}


\newtext{
The main objective is related to the numerous maneuvers that must be performed by the operators.
Due to space restriction, they are not all enumerated, but two examples are provided.
First, the solver is led to minimize the number of stops and starts of the hydro generators on the time horizon:
}

\footnotesize
\begin{equation}
\min \sum_{\ixCfg \in \sCfg} \sum_{\ixT \in \sTf}
  \costManCfgT{\ixCfg}{\ixT}|\varACfgT{\ixCfg}{\ixT} - \varACfgT{\ixCfg}{\ixT - 1}|
\end{equation}
\normalsize

\newtext{
Another objective --- written here in compact form, but linearized in the mathematical model --- aims to minimize the setpoint changes of generators that are not participating to the economic dispatch when there is no starting or stopping of unit in the hydro plant:
}

\footnotesize
\begin{equation}\label{obj:setpoints}
\min \sum_{\ixG \in \sGNed} \sum_{\ixT \in \sTf}
  \costSetGT{\ixG}{\ixT}|\varPGT{\ixG}{\ixT} - \varPGT{\ixG}{\ixT - 1}| \bigwedge_{\ixG' \in \sGHp{\mHpG{\ixG}}} (\varAGT{\ixG'}{\ixT} = \varAGT{\ixG'}{\ixT - 1})
\end{equation}
\normalsize

%
%

\newtext{
Several other secondary objective -- that can optionally be enabled or disabled by the operators -- are introduced to provide some control in case of specific scenarios.
For instance, an objective aims to minimize the gap between the real and optimal yield of each hydro generator at any time step.
Finally, all those different objectives are brought to the same scale with cost coefficients that need to be balanced to be consistent with the desired result.
}

\subsection{Implementation status and challenges}\label{ss:report}


\begin{figure}
  \centering
  \includegraphics[width=.85\linewidth]{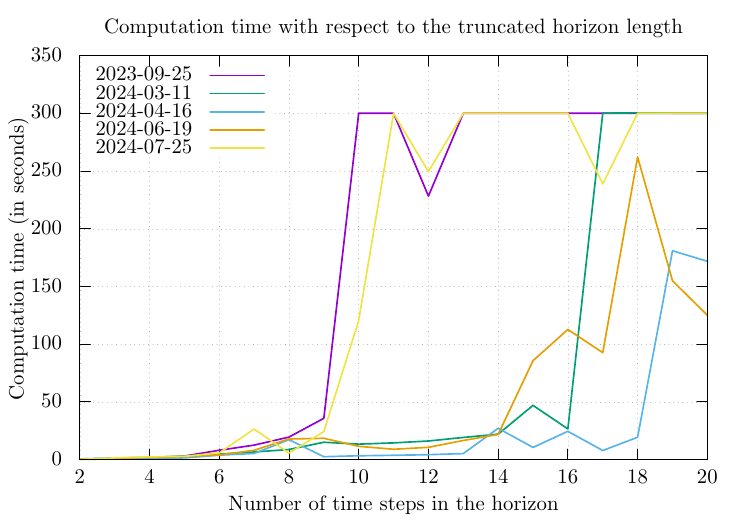}
  \caption{Computation time for solving some real HQ instances with default settings. Timeout was fixed at 300 seconds for each time step.}\label{fig:computation-time}
\end{figure}

\begin{figure}
  \centering
  \includegraphics[width=.85\linewidth]{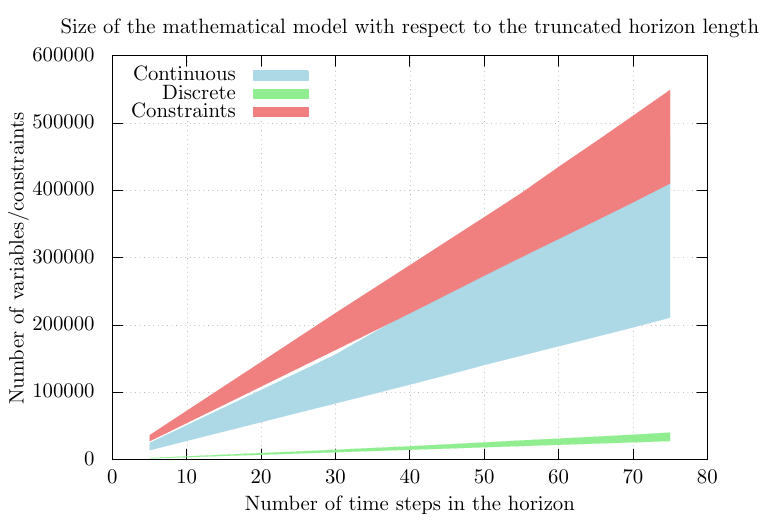}
  \caption{Number of continuous variables, binary variables and constraints of the mathematical problems for the five studied datasets.}\label{fig:model-size}
\end{figure}

\newtext{
Figure~\ref{fig:computation-time} gives an overview of preliminary attempts to
solve five real HQ instances, identified by their date.
In each case, the instances were obtained by truncation, keeping only the $n$ first time steps, making $n$ vary between 2 and 20.
A $300$ seconds timeout was fixed.
A first notable observation is that there is a significant volatility of time resolution with respect to the considered scenarios. Two out of five datasets quickly reach the $300$ seconds barrier, while two other datasets do not.
Also, the curves are not always increasing. In some cases, more data can lead CPLEX to a speed up in its solving time.
Figure~\ref{fig:model-size} shows the variation in the model sizes for the 5 datasets, where the three 2D cones illustrate the minimum and maximum number of continuous variables, discrete variables and constraints.
For instance, when taking 75 time steps horizons, the number of discrete variables is around 30,000, the number of continuous variables is roughly between 200,000 and 400,000, and the number of constraints is roughly between 400,000 and 550,000 for each of the five datasets.
These variations in size partially explain the volatility observed in the resolution times of Figure~\ref{fig:computation-time}.
}

\newtext{
For operational consideration, the ideal objective would be to solve the HTSCUC problem on a rolling window in less than 10 minutes to do real-time optimization of the water resources.
Moreover, to prevent boundary conditions problems at the end of the horizon, the aim is to solve the problem on a 48 hours horizon, so that the first 24 hours can be relied upon, while the last 24 hours can be monitored to ensure that the provided solution makes sense.

The current version of RALPH is a hybrid between human input and optimization tool.
An operator manually inputs preferred configurations for each plant based on a guideline produced by the generation owner (GO).
The optimization calculates all intermediary variables like power flow, water flow or generation by plant.
The operators then need to verify if any constraints like stability limits are exceeded.
If so, they have to manually change the configuration from the guideline until the unit commitment solution respects all reliability constraints.
This process takes hours and can only be done once during the day-ahead planning period.
The new RALPH model presented in this paper seeks to fully automate the solving process by adding the configuration model and the stability limit constraints.
It will} provide a flexible tool to the dispatchers, allowing them to react to different events, such as renewable energy forecast errors or important generation losses that can happen during real-time operations.
It also needs to give a robust solution that does not fluctuate when only minor input changes happen.
To achieve this ambitious goal, Hydro-Quebec is currently developing its software in C++ with a focus on computation performance.
It is also exploring different solutions for the resolution like temporal relaxation and new linearization approaches for the stability limits that would reduce the use of binary variables.



\section{Concluding Remarks}
In this paper, generation operation and planning under stability constraints at Hydro-Quebec were discussed. 
First, the methodology to transform stability limits obtained from transient stability studies into MILP constraints was presented. 
Secondly, the model used for reserve monitoring and reserve restoration used by ALFRED was described. 
Finally, HTSCUC model was also detailed.
While ALFRED is already used by system operators, outage coordinators, and operation engineers, the HTSCUC problem is hard to solve and further research is required to bring solving time into an acceptable timeframe in order to be beneficial to the system operators. \newtext{While the tools described are built to answer specific HQ needs, the rapid rise in inverter based resources (IBR) all over the world will force many operators to address stability issues into their day to day planning. This paper demonstrates that it is possible to consider transient stability problems inside traditionnal MILP problems and can be a starting point for utilities wanting to model stability constraints caused by IBR.}


\bibliographystyle{IEEEtran}
\bibliography{Reference}


 




\vfill

\end{document}